\documentclass[prd,showpacs]{revtex4}

\usepackage{amsmath,amsfonts}
\usepackage[dvips]{graphicx}

\newtheorem{theorem}{Theorem}
\newcommand{\re}{\mathop{\mathrm{Re}}}
\newcommand{\tr}{\mathop{\mathrm{tr}}}

\begin{document}

\title{Dynamical system approach to cosmological models with 
a varying speed of light} 

\author{Marek Szyd{\l}owski}
\email{uoszydlo@cyf-kr.edu.pl}
\affiliation{Astronomical Observatory, Jagiellonian University \\
Orla 171, 30-244 Krak{\'o}w, Poland} 

\author{Adam Krawiec}
\email{uukrawie@cyf-kr.edu.pl}
\affiliation{Institute of Public Affairs, Jagiellonian University \\
Rynek G{\l}{\'o}wny 8, 31-042 Krak{\'o}w, Poland}

%\date{}

\begin{abstract}
Methods of dynamical systems have been used to study homogeneous and 
isotropic cosmological models with a varying speed of light (VSL). 
We propose two methods of reduction of dynamics to the form of planar 
Hamiltonian dynamical systems for models with a time dependent equation 
of state. The solutions are analyzed on two-dimensional phase space in 
the variables $(x, \dot{x})$ where $x$ is a function of a scale factor $a$. 
Then we show how the horizon problem may be solved on some 
evolutional paths. It is shown that the models with negative 
curvature overcome the horizon and flatness problems. The presented 
method of reduction can be adopted to the analysis of dynamics of 
the universe with the general form of the equation of state 
$p=\gamma(a)\epsilon$. This is demonstrated using as an example the 
dynamics of VSL models filled with a non-interacting fluid. We demonstrate 
a new type of evolution near the initial singularity caused by a varying 
speed of light. The singularity-free oscillating universes are also 
admitted for positive cosmological constant. We consider a quantum 
VSL FRW closed model with radiation and show that the highest tunnelling 
rate occurs for a constant velocity of light if $c(a) \propto a^n$ and 
$-1 < n \le 0$. It is also proved that the considered class of models 
is structurally unstable for the case of $n<0$.
\end{abstract}


\maketitle

\section{Introduction}

Although the standard cosmological model is usually believed to 
be a correct picture of our universe \cite{Kolb90} it has still 
some difficulties among which the flatness and horizon problems are 
most widely known. The existence of the large-scale structure in the 
universe extending to the limit of the deepest surveys is another 
mystery. Its very presence implies the appearance of some seeds for 
this structure in the early universe. In the standard big-bang 
scenario they should be built in, which is rather an undesired 
feature of the theory. Therefore, so great attention has been paid 
to inflationary universe models which, albeit invoking an exotic (if not 
hypothetical) physics, were able to provide at least some hope for 
a consistent explanation of both flatness and horizon problems 
as well as the origin of seeds for the large-scale structure. 
The results of the early universe physics lead us to expect 
the occurrence of phase transitions when the universe was young, 
hot, and dense. 

The varying speed of light (VSL) cosmology, seen as an alternative to the 
inflation theory, was proposed by Moffat \cite{Moffat93a,Moffat93b} who 
conjectured that a spontaneous breaking of the local Lorentz invariance and 
diffeomorphism invariance associated with a first order phase-transition 
can lead to the variation of the speed of light in the early universe. 
This idea was revived by Albrecht and Magueijo \cite{Albrecht99} and was 
given further consideration by Barrow \cite{Barrow99b,Barrow00}. Barrow 
showed that the conception of VSL can lead to the solution of flatness, 
horizon, and monopole problems if the speed of light falls at an appropriate 
rate. The dynamics of VSL was widely studied in theoretical as well as 
empirical contexts \cite{Albrecht99b,Alexander99,Avelino99,Avelino00,%
Barrow98,Barrow99,Bassett00,Clayton99,Drummond99,Moffat98}. 

The main motivation for the study of VSL models is to seek explanations 
for some unusual properties of the universe and to overcome some of 
the shortcomings of inflation scenario \cite{Barrow02}. In particular, 
there is an empirical evidence of varying with time the fine structure 
constant in the context of the consistency of quasar absorption spectra 
\cite{Webb99}. Moreover, unlike the inflation the VSL theory provides 
a solution of the cosmological constant problem. However, it cannot 
solve the isotropy problem. It is also interesting to evaluate the power 
of this model to explain the acceleration problem 
\cite{Perlmutter98,Schmidt98}.

Of course, the VSL model as well as other models discussed in the literature 
have an ad hoc element (variable $c$) not yet firmly founded within any 
existing physical theories. This feature does not seems to be so exotic to 
discard these models from discussion in the scientific community. Some 
brilliant arguments justifying this approach are given by Albrecht and 
Magueijo \cite{Albrecht99}. 

The varying speed of light models which provide decent fits to 
the real universe are characterized by the speed of light or gravitational 
coupling which varies with time in the very early universe but is nearly 
constant today. Because there are stringent bounds on how fast these 
constants can vary with time after the first few seconds, the models 
which dynamics we study in the paper are only relevant in the very early 
universe. This should be made very clear by using the phase space approach 
and its tools for classifying the qualitative types of solutions 
\cite{Bogoyavlensky85, Szydlowski90}. 

The present paper is a continuation of previous papers 
\cite{Biesiada00,Biesiada03} 
on the dynamics of VSL cosmology. We introduce a simple framework 
which allows to study the dynamics of VSL models in a general way, 
independent of any specific assumption about the equation of state, or
the behaviour of the scale factor $a(t)$ near the spatial singularity. 
We formulate a VSL Friedmann-Robertson-Walker cosmology as 
a two-dimensional dynamical system and we discuss its properties on phase 
portraits where trajectories represent all solutions for all physically 
admissible initial conditions. The methods of dynamical systems allow to 
indicate how the existence of certain desired physical effects depends on 
the choice of initial conditions and to analyse how these initial conditions 
determining the corresponding solutions are distributed in the phase space. 

Our main goal was to perform a global analysis of the dynamics of VSL 
cosmological models. We avoid the assumption of power type evolution 
in the VSL models, which are represented by critical points (singular 
solutions) in the phase space. We analyse the dynamics of the models 
on the phase plane and discuss how different trajectories representing 
non-singular solutions can solve cosmological puzzles. We conclude that  
models with the negative curvature and positive cosmological constant are 
preferred (in the sense that they have the largest set of initial 
conditions leading to the solution of flatness and horizon problems). 

On the other hand we present two arguments which distinguish the FRW 
models with constant velocity of light. The theoretical one is that 
the VSL FRW models are structurally unstable if $c(a) \propto a^n$ 
and $n<0$ contrary to classical FRW models. The quantum mechanical one 
is that if a closed universe was born from a quantum fluctuation via 
the quantum tunnelling process then the most probable universe is that 
with $c = \text{const}$. In this interval the potential function 
preserves its classical character and the universe tunnels from a zero size. 

The dynamics of considered cosmological models is reduced to the dynamics 
of a unit mass particle in one-dimensional potential. Then different 
physical properties like flatness, horizon and cosmological contant 
problems can be formulated in terms of the diagram of potential function 
of the system. 

\section{The method of the dynamical system stability}

First of all, equations describing a cosmological model should be 
reduced to the form of a dynamical system 
\[
\dot{x}_{i} = \frac{dx_{i}}{dt} = f_{i}(x_{1},\ldots,x_{n}), 
\quad i=1,\ldots,n
\]
in such a way that the solution with a static microspace (or 
other solutions of interest) is a critical point of the system 
$(x_{1}^{*},\ldots,x_{n}^{*})$, i.e.\ for every $i$, 
$f_{i}(x_{1}^{*},\ldots,x_{n}^{*})=0$ $(i=1,\ldots, n)$.

If a critical point is non-degenerate, i.e., at this point 
all real parts of the eigenvalues $\re \lambda_{i}$ 
of the linearization matrix 
\[
A_{j}^{i} = \left. \frac{\partial f_{i}}{\partial x_{j}} 
\right|_{x=x^{*}}
\]
do not vanish, then there is a one-to-one continuous 
mapping of a neighbourhood of this point which transforms 
trajectories of the original system into trajectories of 
the linearized system 
\[
\frac{dx_{i}}{dt} = \sum \frac{\partial f_{i}}{\partial x_{j}}(x^{*}) 
(x_{j} - x_{j}^{*}).
\]
In this sense, the qualitative behaviour of the original system 
is equivalent to the behaviour of its linearized part. 
If $(\xi_{i}^{1},\ldots,\xi_{i}^{n})$ are eigenvectors of 
the linearization matrix $A_{j}^{i}$, the solution of 
the linearized system has, in general, the following form
\[
x_{i}(t) - x_{i}^{*} = \re \sum_{k=1}^{n} 
C_{k} \xi_{i}^{k} e^{\lambda_{k}t}
\]
where $C_{k}$ are constants. A non-degenerate critical point 
is called an attracting point if, for all eigenvalues, 
$\re \lambda_{i} < 0$. In this case, all trajectories 
from a neighbourhood of this point go to it when $t \to \infty$.
A non-degenerate critical point is called an repulsing point if, 
for all eigenvalues, $\re \lambda_{i} > 0$. In this case, 
all trajectories from a neighbourhood of this point go to it when 
$t \to - \infty$. A non-degenerate critical point is said to be 
an unstable saddle point and if a dynamical system has, 
at $(x_{1}^{*},\ldots,x_{n}^{*})$, $d$ eigenvalues with positive 
real parts and $n-d$ eigenvalues with negative real parts 
$(d=1,\ldots,n-1)$. 

When investigating the stability of solutions with a static 
microspace the following theorem proves to be of special interest. 
If $x^{*}$ is a non-degenerate critical point and if the dynamical 
system has, at $x^{*}$, $d$ eigenvalues $\lambda_{1},\dots,\lambda_{d}$ 
with negative real parts, then there exists (locally) an invariant 
$d$-dimensional stable manifold $W^{d}_{\text{st}}$, on which all 
trajectories of the system go to $x^{*}$ as $t \to \infty$. 
(A manifold $M$ is said to be an invariant manifold of a system 
if every trajectory passing through a non-degenerate point of $M$ 
lies entirely in $M$ (for $-\infty < t < +\infty$). For every such
solution, there exists the asymptotic
\begin{equation}
\lim_{t \to \infty} t^{-1} \ln \left\{ \sum_{j=1}^{n} 
[ x_{j}(t) - x_{j}^{*} ]^{2} \right\}^{1/2} = \alpha_{i}
\end{equation}
for a certain $i$. Similarly, if at the critical point $x^{*}$ 
the system has $k$ eigenvalues with positive real parts then 
there exists an invariant $k$-dimensional unstable manifold 
$W^{k}_{\text{unst}}$ on which all trajectories go away 
from the critical point \cite{Bogoyavlensky85}. 

From the above theorem follows that, for a saddle point, 
there are two invariant manifolds $W^{d}_{\text{st}}$ 
and $W^{n-d}_{\text{unst}}$ containing this point and 
filled with trajectories (separatrices) going to and going 
away from the critical point. All other trajectories (not 
contained in $W^{d}_{\text{st}}$ or in 
$W^{n-d}_{\text{unst}}$) do not meet the critical point 
in question. 

For the complete construction of phase portraits in a plane 
it is necessary to know how trajectories of dynamical system 
behave at infinity. Let us take as an example the two-dimensional 
system
\begin{align}
\dot{x} &= P(x,y) \\
\dot{y} &= Q(x,y).
\end{align}
In the case of polynomial right-hand sides 
one usually introduces projective coordinates, eg. $z=1/x$, 
$u=y/x$ or $v=1/y$, $w=x/y$. Two maps $(z,u)$ and $(v,w)$ 
are equivalent if and only if $u \neq 0$ and $v \neq 0$. Infinitely 
distant points of $(x,y)$-plane correspond to a circle $S^{1}$ 
which can be covered by two lines $z=0$, $-\infty < u < \infty$ 
and $w=0$, $-\infty < v < \infty$. The original system in the 
projective coordinates $(z,u)$ and after the time reparametrization 
$\tau \to \tau_{1} \colon d\tau_{1} = xd\tau$ assumes the form 
\begin{align}
\dot{z} &= z P^{*}(z,u) \\
\dot{u} &= Q^{*}(z,u) - u P^{*}(z,u)
\end{align}
where
\begin{align*}
P^{*}(z,u) &= z^{2} P(1/z,u/z) \\
Q^{*}(z,u) &= z^{2} Q(1/z,u/z)
\end{align*}
and dot denotes differentiation with respect to new time $\tau_{1}$.

On a similar way, in the projective coordinates $(v,w)$ and in the new time 
$\tau_{2} \colon d\tau_{2} = yd\tau$ we obtain 
\begin{align}
\dot{v} &= - v Q^{*}(v,w) \\
\dot{w} &= P^{*}(v,w) - w Q^{*}(v,w)
\end{align}
where
\begin{align*}
P^{*}(v,w) &= v^{2} P(1/v,w/v) \\
Q^{*}(v,w) &= v^{2} Q(1/v,w/v)
\end{align*}
and dot denotes differentiation with respect to time $\tau_{2}$.

The idea of structural stability was introduced by Andronov and Pontryagin
\cite{Andronov37}. A dynamical system $S$ is said to be structurally stable
if there exist dynamical systems in the space of all dynamical systems 
which are close, in the metric sense, to $S$ or are topologically equivalent 
to $S$. Instead of finding and analyzing an individual solution of a model,
a space of all possible solutions is investigated. A given physical property 
is believed to be `realistic' if it can be attributed to large subsets of 
models within a space of all possible solutions or if it possesses a certain 
stability, i.e., if it is also shared by a slightly perturbed model. There is 
a well established opinion among specialists that realistic models should 
be structurally stable. What does the structural stability mean 
in physics? The problem is in principle open in higher than 2-dimensional 
case where according to Smale there are large subsets of structurally 
unstable systems in the space of all dynamical systems \cite{Smale80}.
For 2-dimensional dynamical systems, as in the considered case, the
Peixoto's theorem says that structurally stable dynamical systems on
compact manifolds form open and dense subsets in the space of all
dynamical systems on the plane. Therefore, it is reasonable to require
the model of a real 2-dimensional problem to be structurally stable.

In our further considerations we will investigate the dynamics of 
dynamical systems in a finite domain of phase space as well as 
at infinity. At this point we would like to recommend 
the presentation of the actual state of the art in the field of 
application of the dynamical systems to general relativity 
\cite{Wainwright97}.

\section{Basic equations of the theory}

Albrecht and Magueijo \cite{Albrecht99} and Barrow \cite{Barrow99b} 
set up a useful framework to discuss the VSL models assuming that 
the time variable $c$ should not introduce changes in the curvature 
terms of the gravitational field equations and that the Einstein 
equations must hold. Because varying $c$ breaks the Lorentz invariance 
the VSL cosmology requires a specific reference frame (including 
a specific choice of a time coordinate) in which changes in the field 
equations are minimal and one postulates it to coincide with the 
cosmological comoving frame. 

In the case of the VSL version of the FRW models (with $\Lambda=0$) 
the scale factor obeys the following dynamical equations 
\begin{gather}
\label{eq:1}
\left( \frac{\dot{a}}{a} \right)^{2} = 
\frac{8\pi G(t)\rho}{3} - \frac{Kc^{2}(t)}{a^{2}(t)} \\ 
\label{eq:2}
\frac{\ddot{a}(t)}{a} = - \frac{4\pi G(t)}{3} 
\left( \rho + \frac{3p}{c^{2}(t)} \right).
\end{gather}
Equation (\ref{eq:2}) is called the Raychaudhuri equation 
and from the above system one can build a generalized 
conservation equation
\begin {equation}
\label{eq:3}
\dot{\rho} + 3 \frac{\dot{a}}{a} \left( \rho + 
\frac{p}{c^{2}(t)} \right) = - \rho \frac{\dot{G}}{G} 
+ \frac{3Kc^{2}}{8\pi G a^{2}} \frac{\dot{c}}{c} 
\end{equation}
in which time dependence of fundamental constants was taken into account 
explicitly. Alternatively one can think of the Raychaudhuri equation 
together with the generalized conservation equation as of a fundamental 
system to which equation~(\ref{eq:1}) is a first integral. 

The fundamental difficulty concerning system~(\ref{eq:1})--(\ref{eq:3}) 
is that it is a non-autonomous system with unknown functions $G(t)$ and 
$c(t)$. In order to be specific in the further analysis we adopt 
Barrow's power-law Ansatz
\begin{equation}
\label{eq:4}
G(t) = G_{0} a(t)^{q} , \quad c(t) = c_{0} a(t)^{n}.
\end{equation}
Moreover, we assume the hydrodynamical energy-momentum tensor 
with the equation of state for the non-interacting multifluid 
\begin{equation}
\label{eq:5}
p = \frac{\sum_{i=1}^{l} \gamma_{i} \rho_{i} c^{2}}{\sum_{i=1}^{l} 
\rho_{i} c^{2}} \epsilon = \gamma(a) \rho c^{2}
\end{equation}
where $\rho_{i} = \rho_{i0} a^{-3(\gamma_{i}+1)}$, and energy density 
$\epsilon = \rho c^2$.

In the special case of the matter and radiation mixture, 
the factor $\gamma(a)$ depending on the scale factor $a$ takes 
the form 
\begin{equation}
\label{eq:6}
\gamma(a) = \frac{1}{3} \frac{1}{\alpha a + 1} , \quad 
p = 0 + \frac{1}{3}\epsilon_{\text{r}} , \quad 
\epsilon = \epsilon_{\text{m}} + \epsilon_{\text{r}}
\end{equation}
where $\alpha = \rho_{{\text{m}} 0} / \rho_{{\text{r}} 0}$.

If we substitute $l=1$ into~(\ref{eq:5}) then we obtain models filled 
with single matter and with the equation of state $p=\gamma \rho c^{2}$. 
Generally, the equations of state for non-interacting fluids with pressure 
$p = \sum_{i} \gamma_{i} \rho_{i} c^{2}$, the equation of state assumes 
the form $p = \gamma(a) \rho c^{2}$ where the factor $\gamma$ can be 
parametrized by the scale factor. This fact is crucial for the reduction 
procedure. 

Power-law Ansatz~(\ref{eq:4}) turns the field equations 
back into an autonomous system. Now we can think about extensions 
of the baseline equations. One can in a straightforward way 
include the cosmological constant $\Lambda$ by introducing 
pressure $p_{\Lambda}$ and energy density $\rho_{\Lambda}$ 
\begin{align}
\label{eq:7}
p_{\Lambda} &= - \rho_{\Lambda} c^{2}(t) \\
\label{eq:8}
\rho_{\Lambda} &= \frac{\Lambda c^{2}(t)}{8 \pi G(t)}.
\end{align}
System (\ref{eq:1})--(\ref{eq:3}) with the cosmological constant 
can be cast into form 
\begin{gather}
\label{eq:9}
\left( \frac{\dot{a}}{a} \right)^{2} = \frac{8 \pi G(t) \rho}{3} 
- \frac{K c^{2}(t)}{a^{2}(t)} + \frac{\Lambda c^{2}(t)}{3} \\
\label{eq:10}
\frac{\ddot{a}}{a} = - \frac{4 \pi G(t)}{3} 
\left( \rho  + \frac{3p}{c^{2}(t)} \right) + \frac{\Lambda c^{2}(t)}{3} \\ 
\label{eq:11}
\dot{\rho} + 3 \frac{\dot{a}}{a} \left( \rho + 
\frac{p}{c^{2}}{c^{2}(t)} \right) = 
- \rho \frac{\dot{G}}{G} + \frac{3Kc^{2}}{8 \pi G a^{2}} 
\frac{\dot{a}}{a}.
\end{gather}
Equation (\ref{eq:11}) is easy to solve only for the case of 
$c=\text{const}$. Therefore, in our consideration of the general 
formulation of dynamics we cannot use an explicit form of a solution 
of equation~(\ref{eq:11}). To avoid this difficulty we consider 
a special procedure of reduction.

\section{Reduction to a planar Hamiltonian system}

\subsection{The general solution of a dynamical problem}

To construct a dynamical system we assume the form of the equation 
of state $p = \gamma(a) \rho c^{2}$ and calculate the density $\rho$ 
using both equation~(\ref{eq:8}) and (\ref{eq:9}). 
For simplicity we focus our attention on the case of $\Sigma = 0$ 
corresponding to the VSL model with matter in the multifluid form. 
Then we obtain from equation~(\ref{eq:9}) 
\begin{equation}
\label{eq:13}
\frac{8 \pi G \rho}{3} = \frac{\dot{a}^{2}}{a^{2}} +
\frac{K c^{2}(t)}{a^{2}} - \frac{\Lambda c^{2}}{3} 
\end{equation}
and from equation~(\ref{eq:10})
\begin{equation}
\label{eq:14}
- \frac{8 \pi G \rho}{3} = \frac{1}{1 + 3 \gamma(a)} 
\left( \frac{2\ddot{a}}{a} - \frac{2\Lambda c^2}{3} \right).
\end{equation}
By adding the sides of above equations we obtain a second order nonlinear 
equation with respect to variable $a$ 
\begin{equation}
\label{eq:15} 
\ddot{a} + \psi(a) \dot{a}^{2} + \kappa(a) = 0
\end{equation}
where
\begin{align*}
\psi(a) &= \frac{1 + 3 \gamma(a)}{2a} \\
\kappa(a) &= \left[ \frac{K}{2a}(1+3\gamma(a)) - 
\frac{\Lambda}{2} a (1 + \gamma(a)) \right] c^{2}(a)
\end{align*}
and only the term $\kappa(a)$ depends on the cosmological constant.
 
Equation~(\ref{eq:15}) can be rewritten as an autonomous dynamical system 
\begin{align}
\label{eq:16}
\dot{a} &= p \\
\label{eq:17}
\dot{p} &= - \psi(a) p^{2} - \kappa(a).
\end{align}
To apply the dynamical system theory, it is useful to reduce system 
(\ref{eq:16})--(\ref{eq:17}) to the form with polynomial right-hand sides 
\begin{align}
\label{eq:18}
a' &= \frac{d a}{d \eta} = p a \\
\label{eq:19}
p' &= \frac{d p}{d \eta} = - \frac{1}{2} (3\gamma(a)+1) p^{2} 
- \phi(a)
\end{align}
where $\phi(a) = a \kappa(a)$ and $t \to \eta \colon \frac{dt}{a} = d\eta$.
The solution of equations~(\ref{eq:18})--(\ref{eq:19}) represents a phase 
curve in the phase space $(a,p)=(a,\dot{a})$. 

The solution of equation~(\ref{eq:15}) may be given after substitution 
$\dot{a} = p(a)$ or equivalently by taking the quotient of 
equations~(\ref{eq:16}) and (\ref{eq:17}). Then we obtain 
\begin{equation}
\label{eq:20}
p \frac{dp}{da} + \psi(a) p^{2} + \kappa(a) = 0.
\end{equation}
Equation~(\ref{eq:20}) takes the form of the Bernoulli equation and 
after standard substitution $u(a) = p^{2}$, $u' = du/da$ we obtain the 
non-autonomous system 
\begin{equation}
\label{eq:21}
u' + 2 \psi(a) u + 2 \kappa(a) = 0.
\end{equation}
Finally, the solution of equation~(\ref{eq:15}), passing through 
the point $(a_{0}, p_{0}(a_{0}))$, can be given in the following 
form 
\[
u(a) = \exp \left( -2 \int_{a_{0}}^{a} \psi(a)da \right) 
\left[ p_{0}^{2}(a_{0}) - \int_{a_{0}}^{a} 2\kappa(a) 
\exp \left( 2 \int_{a_{0}}^{a} \psi(a)da \right) da \right]
\]
i.e.
\[
p^{2}(a) = \exp \left( - \int_{a_{0}}^{a} \frac{3\gamma(a)+1}{a} da \right) 
\left\{ p_{0}^{2}(a_{0}) 
-2 \int_{a_{0}}^{a} \left[ \frac{K(1+3\gamma(a))}{2a} - 
\Lambda a (\gamma(a)+1) \right] \exp \left( \int_{a_{0}}^{a} 
\frac{3\gamma(a)+1}{a} da \right) da \right\}.
\]

To consider the case of a mixture of matter and radiation, we 
substitute the special form of $\gamma(a)$ from formula~(\ref{eq:6}) 
and then we obtain 
\begin{equation}
\label{eq:22}
p^{2}(a) =  \frac{a_{0}^{2}(\alpha a+1)}{a^{2}(\alpha a_{0}+1)} 
\left\{ p_{0}^{2}(a_{0}) 
- 2 \int_{a_{0}}^{a} \frac{a^{2}}{\alpha a + 1} da 
\left[ \frac{K(1+3\gamma(a))}{2a} - \Lambda a 
\left( 1+\frac{1}{3(\alpha a+1)} \right) \right] c^{2}(a) \right\}
\end{equation}
and the general solution of equation~(\ref{eq:21}) has the form 
\begin{equation}
\label{eq:23}
u(a) = \exp \left( -2 \int_{a_{0}}^{a} \psi(a)da \right) \left[ C 
- 2 \int_{a_{0}}^{a} \kappa(a) \exp \left( 2\int_{a_{0}}^{a} \psi(a)da 
\right) da \right].
\end{equation}
It means that the following expression can be treated as a first 
integral of system~(\ref{eq:18})--(\ref{eq:19}). It is characteristic 
for dynamical systems of general relativity and cosmology that a 
first integral can be used in constructing a Hamiltonian function. 
A first integral can be represented as algebraic curves in the phase 
space. These algebraic curves are given by 
\begin{equation}
\label{eq:24}
p^{2}(a) \exp \left( 2 \int_{a_{0}}^{a} \psi(a)da \right) 
+ 2 \int_{a_{0}}^{a} \kappa(a) \exp \left( 2\int_{a_{0}}^{a} \psi(a)da 
\right) da = C.
\end{equation}
And then in the considered case we obtain 
\begin{equation}
\label{eq:25}
\dot{a}^{2} \frac{a^{2}}{\alpha a + 1} + 2 \int_{a_0}^{a} \kappa(a)
\frac{a^{2}}{\alpha a + 1} da = C
\end{equation}
Now if we introduce a new variable, $x$, such that 
\[
\frac{1}{\sqrt{2}} dx = \frac{ada}{\sqrt{\alpha a+1}}
\]
the above relation can be written as 
\[
\frac{\dot{x}^{2}}{2} + V(x) = V(a_{0}) = \text{const} 
\]
where
\begin{equation}
\label{eq:26}
V(x) = 2 \int^{x} \frac{\kappa(a(x)) a(x)}{\alpha a(x) +1} dx 
\end{equation}
plays the role of the potential 
with $a(x) \colon \frac{1}{\sqrt{2}}x = \frac{2}{3\alpha^2} 
\sqrt{\alpha a+1} (\alpha a -2)$.

This procedure successfully works for any function $\psi(a)$. It is 
sufficient to replace the expression $\frac{a^{2}}{\alpha a+1}$ in 
the considered case by $\exp(2\int^{a} \psi(a)da) \equiv \phi^{2}$.

In the special case of $\alpha = 0$ we obtain that 
$p=\frac{1}{3}\epsilon$ and we can use the standard formalism 
considered in \cite{Szydlowski84}. 

Now we can see from formula~(\ref{eq:22}) that an algebraic 
curves on which lie trajectories of the system take the 
complicated form. Therefore it is useful to visualize them 
in the phase space. Using the form of the above first integral 
we can also classify all possible solutions by considering 
a limiting curve $\dot{x} = 0$ and derive the relation 
$\Lambda(a)$ as was presented in the classical case \cite{Szydlowski84}.

There are two kinds of the different methods of reducing 
equation~(\ref{eq:15}) to the form of Newtonian equation of motion in 
a one-dimensional configuration space. First, after introducing the new 
rescaled variable $a \to x$, we obtain dynamics in the form 
$\ddot{x} = - \frac{\partial V}{\partial x}$. 
Second, after introducing the new time variable, say $\tau(t)$ 
defined in such a way that the term $\psi(a) \dot{a}^{2}$ can be 
dropped in this parametrization and we obtain dynamics in the form 
$x'' = \frac{d^2 a}{d \tau^2} = - \frac{\partial V}{\partial a}$.

Let us note that the function $V(a(x))$ plays the role of the potential 
for a particle which the position is given by $x$ and motion described by 
\[
\ddot{x} = - \frac{dV(a(x))}{dx}. 
\]
Szyd{\l}owski {\em et al.\ } \cite{Szydlowski84} showed that the choice of 
new variables $x=x(a)$ allows to reduce the dynamics of classical 
FRW models with matter or radiation to a one-dimensional Newtonian 
equation of motion. It is also interesting that there is a similar 
possibility of reducing system (\ref{eq:15}) to the Hamiltonian form 
for any equation of state $\gamma = \gamma(a)$, for example 
for any mixture of non-interacting fluids. To perform this let us consider 
the general nonlinear reparametrization of variable $a$ such that 
\begin{equation}
x = a^{D(\gamma(a))} = a^{D(a)}
\end{equation}
It can be shown that equation (\ref{eq:15}) can be reduced 
to the form of a Newtonian equation of motion, $\ddot{x} =
- dV(x)/dx$, if the multiplicative coefficient appearing 
in $\dot{a}^2$ vanishes. This condition gives us the following 
\begin{equation}
\label{eq:43}
D_{aa} (\ln a) + \frac{2D_{a}}{a} - \frac{D}{a^{2}}
\left( D_{a} \ln a + \frac{D}{a} \right)^{2} = 
\psi(a) \left( D_{a} \ln a + \frac{D}{a} \right) 
\end{equation}
where $D_{a} = \partial D/\partial a$.
In the special case of $\gamma(a) = \text{const}$ ($D_{a}=0$), classical 
results can be recovered \cite{Szydlowski84}; we have $D=2$ for pure radiation 
and $D=3/2$ for dust and $D=\frac{3}{2}(1+\gamma)$ for perfect 
fluid with $p=\gamma \rho$. If $\gamma = \gamma(a)$ 
is any function of $a$ then $D(a)$ is a solution of the above equation. 
After simple substitution 
\begin{equation}
\label{eq:44}
D_{a} (\ln a) + \frac{D(a)}{a} = z(a)
\end{equation}
we obtain that $z(a)$ is a solution of the equation 
\begin{equation}
z + \frac{1}{z} \frac{dz}{da} = \psi(a).
\end{equation}
In this new variable
\[
\ddot{x} = - \kappa(a(x)) z(a(x)) x = - \frac{\partial V}{\partial x}
\]
and the dynamics is reduced to the case of a nonlinear oscillator with 
`spring-like tension' $k(x) = \kappa(a) z(a)$. 

The information about the equation of state is hidden in the function 
$\psi(a)$ and after finding the solution $z(a)$ for a specific form 
of the equation of state it should be easy to find $D(a)$ from equation 
(\ref{eq:44}). It can be easily shown that the corresponding equation 
determining $z(a)$ is the Bernoulli equation for which the solution is 
\[
z(a) = \frac{\phi(a)}{\int^{a} \phi(a) da} = 
\frac{d\phantom{a}}{da} \ln \left( \int^{a} \phi(a) da \right)
\]
where $\phi(a) \equiv \exp(\int \psi(a) da)$. If we put $z(a)$ into 
(\ref{eq:44}) we can find that 
\[
D(a) = \frac{\ln \int^{a} \phi(a) da}{\ln a} = \log_{a} \int^{a} \phi(a)da.
\]
For the case of mixture of radiation and dust it has the simple form 
\[
z(a) = \begin{cases} 
\frac{3\alpha^{2}}{2} \left| \frac{a}{(\alpha a+1)(\alpha a-2)} \right| & 
\text{for } \alpha \neq 0 \\
\frac{3}{2a} & \text{for } \alpha = \infty \\
\frac{2}{a} & \text{for } \alpha = 0 \end{cases}
\]
where $\phi(a) = \frac{a}{\sqrt{\alpha a+1}}$ and 
$\int^{a} \phi(a)da=\frac{2}{3\alpha^{2}} |\sqrt{\alpha a+1} (\alpha a-2)|$. 

Therefore we obtain for the case of non-interacting matter and radiation 
\[
D(a) = \begin{cases}
\log_{a} \left( \frac{2\sqrt{\alpha a+1} |\alpha a-2|}{3\alpha^2}\right) & 
\text{for } \alpha \neq 0 \\
2 & \text{for } \alpha = 0 \\
\frac{3}{2} & \text{for } \alpha = \infty.
\end{cases}
\]
It can be proved that in a general situation we have the following 
relation 
\begin{equation}
\label{eq:45}
\int^{a} \phi(a) da = x(a) = \int^{a} \sqrt{a} 
e^{\left( \frac{3}{2} \int^{a} \frac{\gamma(a')}{a'} da'\right)} da
\end{equation}
and 
\[
x(a) = \begin{cases}
a^{\log_{a}(\frac{2\sqrt{\alpha a+1}|\alpha a-2|}{3\alpha^2})} & 
\text{for } \alpha \neq 0 \\
a^{2} & \text{for } \alpha = 0 \\
a^{3/2} & \text{for } \alpha = \infty 
\end{cases}
\]
i.e., for any fluid (or its mixture) which satisfies the equation of 
state for so-called `quintessence' matter $p=\gamma(a) \rho c^{2}(a)$ 
we can always find the corresponding $D(a)$.

Due to (\ref{eq:45}) the equation of motion can be rewritten to the 
simplest form 
\begin{equation}
\label{eq:46}
\ddot{x} = - \kappa(a(x)) \phi(a(x)), \quad 
V(x) = \int^{x} \kappa(a(x)) \phi(a(x)) dx.
\end{equation}

Let us note that there is also a possibility to generalize such a 
result to the case of non-vanishing shear in B(I) or B(V) models 
when $\sigma \propto x^{3/2}$. 

In the second approach it is useful to reparametrize the time 
variable $t$ in such a way that 
\begin{equation}
t \to \tau \colon dt = \phi(a(\tau))d\tau
\end{equation}
where $\psi(a)$ is a yet-to-be-determined function which should be chosen in 
such a way that term $\psi(a) \dot{a}^2$ is absent in (\ref{eq:15}). 
We can do that provided that $\phi$ satisfies condition 
\begin{equation}
\phi \equiv \exp \int^{a} \psi(a) da.
\end{equation}
Then equation (\ref{eq:15}) assumes the form similar to the 
equation for the motion of a non-relativistic particle in the external 
field with the potential $V(a)$, namely
\begin{equation}
\frac{d^2 a}{d\tau^2} \equiv a'' = - \frac{\partial V}{\partial a}
\end{equation}
and 
\begin{equation}
V(a) = \int^{a} \kappa(a) \phi^{2}(a) da
\end{equation}
where 
\begin{align*}
\psi(a) &= \frac{1+3\gamma(a)}{2a} \\
\kappa(a) &= \left[ \frac{K}{2a} (1+3\gamma(a)) - \frac{\Lambda}{2} 
a(1+\gamma(a)) \right] c^{2}(a).
\end{align*}
Therefore for such a system the Hamiltonian takes the form 
\begin{equation}
{\cal H}(a,p) = \frac{p^2}{2} + V(a) = E = \text{const}
\end{equation}
where the correspondence to the vacuum case is reached after putting 
$E=0$ and $\gamma(a)=0$. 

The advantage of this procedure of reduction is its simplicity. 
The new time variable $\tau$ is a monotonic function of 
Newtonian time $t$ and motion is represented in the form of 
one-dimensional Hamiltonian system with the potential 
\begin{equation}
\label{eq:50}
V(a) = \int^{a} \kappa(a) \phi^{2}(a) da = 
c_{0}^{2} \int^{a} \left[ \frac{K}{2a} (1+3\gamma(a)) - \frac{\Lambda}{2} 
a(1+\gamma(a)) \right] a^{2n} \phi^{2}(a) da.
\end{equation}

By comparison potential (\ref{eq:26}) and potential (\ref{eq:50}) 
we can observe that both procedures give rise to the same form of the 
potential function as a function of $a$. Whereas the second approach 
seems to be simpler, the first one has an advantage that it allows to 
discuss dynamics in the origin time. In both cases we do not explicit 
integrate continuity equation (\ref{eq:11}) which gives us the relation 
$\rho(a)$. The effect of matter content is included in $\rho(a)$ and 
energy constant $E$. 

For the special case of matter content in the form of non-interacting 
matter and radiation we obtain 
\begin{equation}
\label{eq:51}
V(a) = \frac{Kc_{0}^{2}}{2} \int^{a} \frac{(\alpha a+2) 
a^{2n+1}}{(\alpha a+1)^2} da + \frac{\Lambda c_{0}^{2}}{2} 
\int^{a} \frac{(3\alpha a+4) a^{2n+3}}{(\alpha a+1)^2} da.
\end{equation}
where integrals in the above form of the potential can be given 
explicitly. 

In the special case of $\Lambda=n=0$ we have 
\begin{equation}
\label{eq:53}
V(a) = \frac{K a^{4}}{2(\alpha a+1)}.
\end{equation}
Therefore the dynamics is given by the Hamiltonian equations 
\begin{align}
\label{eq:54}
a' &= \frac{\partial \mathcal{H}}{\partial p} \\
\label{eq:55}
p' &= - \frac{\partial \mathcal{H}}{\partial a} 
\end{align}
which constitute the two-dimensional dynamical system. 
Entire evolution is represented by an evolutional path 
on the plane $(a,p)$. The domain of acceleration 
$\ddot{a}(t)>0$ is determined by the condition 
\begin{equation}
\label{eq:56}
a'' - (a')^{2} \psi(a) > 0.
\end{equation}
The above condition can be rewritten as 
\begin{equation}
\label{eq:57}
- \frac{\partial V}{\partial a} - (a')^{2} \psi(a) > 0.
\end{equation}
Let us note that equation (\ref{eq:57}) can be formulated equivalently as a 
following condition in some domain of the configuration space 
$\{ a \colon a \ge 0 \}$ 
\begin{equation}
\label{eq:58}
- \frac{\partial V}{\partial a} - 2(E-V(a)) \psi(a) > 0
\end{equation}
or
\begin{equation}
\label{eq:59}
\frac{1}{\psi(a)} \frac{\partial V}{\partial a} - 2V(a) < -2E
\end{equation}
where $\frac{\partial V}{\partial a} = - \kappa(a) \phi^{2}(a)$ and 
$\psi(a) = \frac{1+3\gamma(a)}{2a}$.

\section{Evolution of the VSL dynamical system on phase diagrams}

\subsection{Background}

In the further qualitative analysis of dynamical system 
(\ref{eq:16})--(\ref{eq:17}) we consider the matter as 
the mixture of radiation and dust. Then system 
(\ref{eq:16})--(\ref{eq:17}) takes the form of the autonomous 
system with rational right-hand sides
\begin{align}
\label{eq:60}
\dot{a} &= p  \\
\label{eq:61}
\dot{p} &= - \frac{\alpha a + 2}{2a(\alpha a + 1)} p^{2} 
- \left( \frac{K(\alpha a + 2)}{2a(\alpha a + 1)} 
- \frac{3 \alpha a + 4}{6(\alpha a+1)} 
\Lambda a \right) c^{2}(a) 
\end{align}
where $c^{2}(a) = c_{0}^{2}a^{2n}$, $n \le 0$, $\alpha = 
\epsilon_{\text{m} 0} / \epsilon_{\text{r} 0}$ 
(for regularization of the system in $a=0$ it is useful as 
in (\ref{eq:18})--(\ref{eq:19}) 
to introduce time $\eta \colon \frac{dt}{a} = d\eta$).  
Of course the above system possesses the first integral in 
form~(\ref{eq:22}). 

In the finite domain, system (\ref{eq:60})--(\ref{eq:61}) has 
at most one critical point which corresponds to an extremum 
of the function $V(a) \colon \left. \frac{dV}{da}\right|_{a=a_{0}} = 0$, 
$p_{0}(a_{0}) = 0$. The stability of this point is determined 
from the convectivity of a diagram of potential function $V(a)$. 
There are two limit cases corresponding to the equation of state: 
of dust ($\alpha \to \infty$) and of pure 
radiation ($\alpha = 0$). From system (\ref{eq:60})--(\ref{eq:61}) 
in the latter case we obtain the VSL system with pure radiation 
\begin{align}
\label{62}
\dot{a} &= p \\
\label{eq:63}
\dot{p} &= - \frac{1}{a} p^{2} - \left( \frac{K}{a} - 
\frac{2}{3} \Lambda a \right) c^{2}(a).
\end{align}

\begin{figure}
\includegraphics[angle=-90,width=0.4\textwidth]{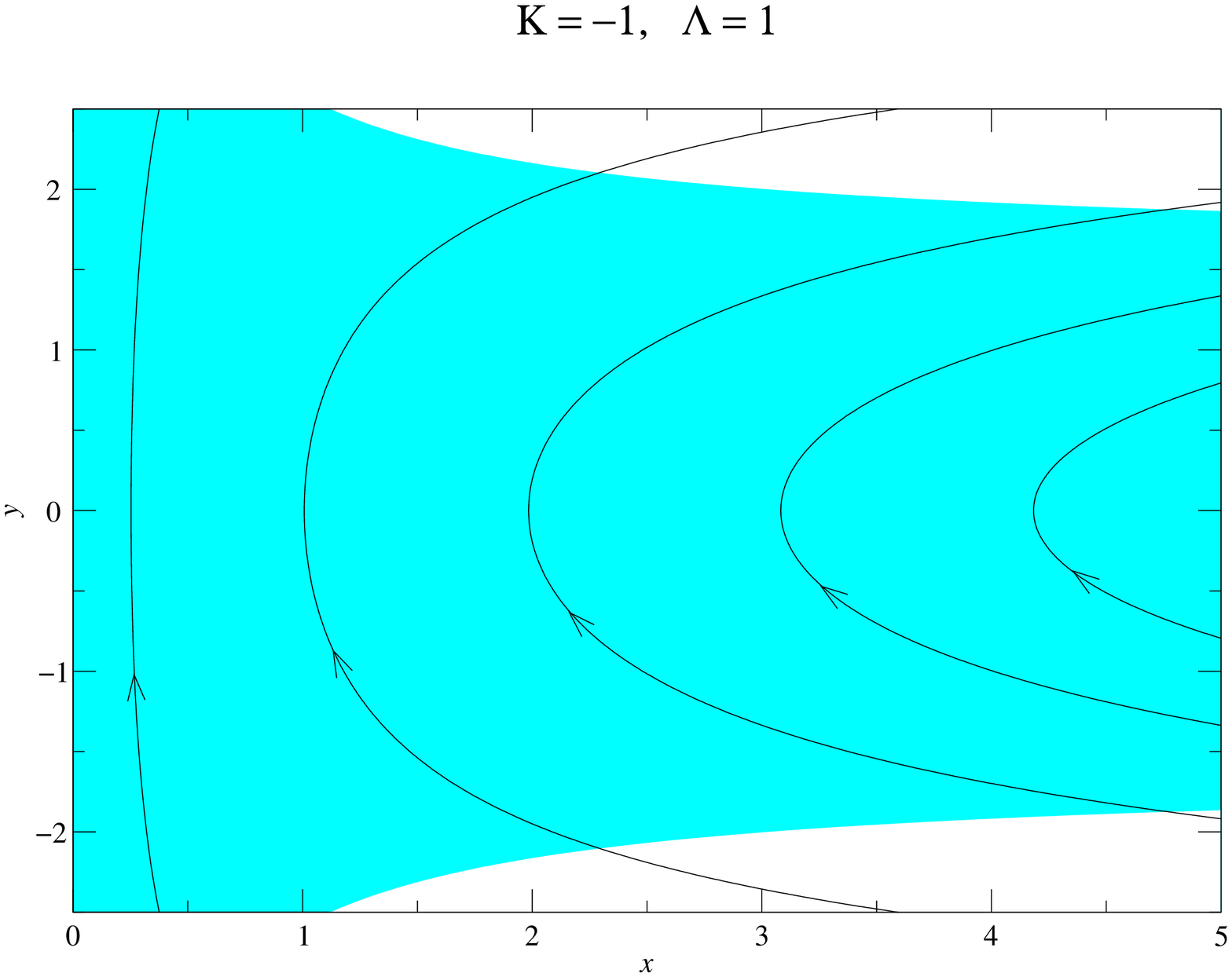}%
\includegraphics[angle=-90,width=0.4\textwidth]{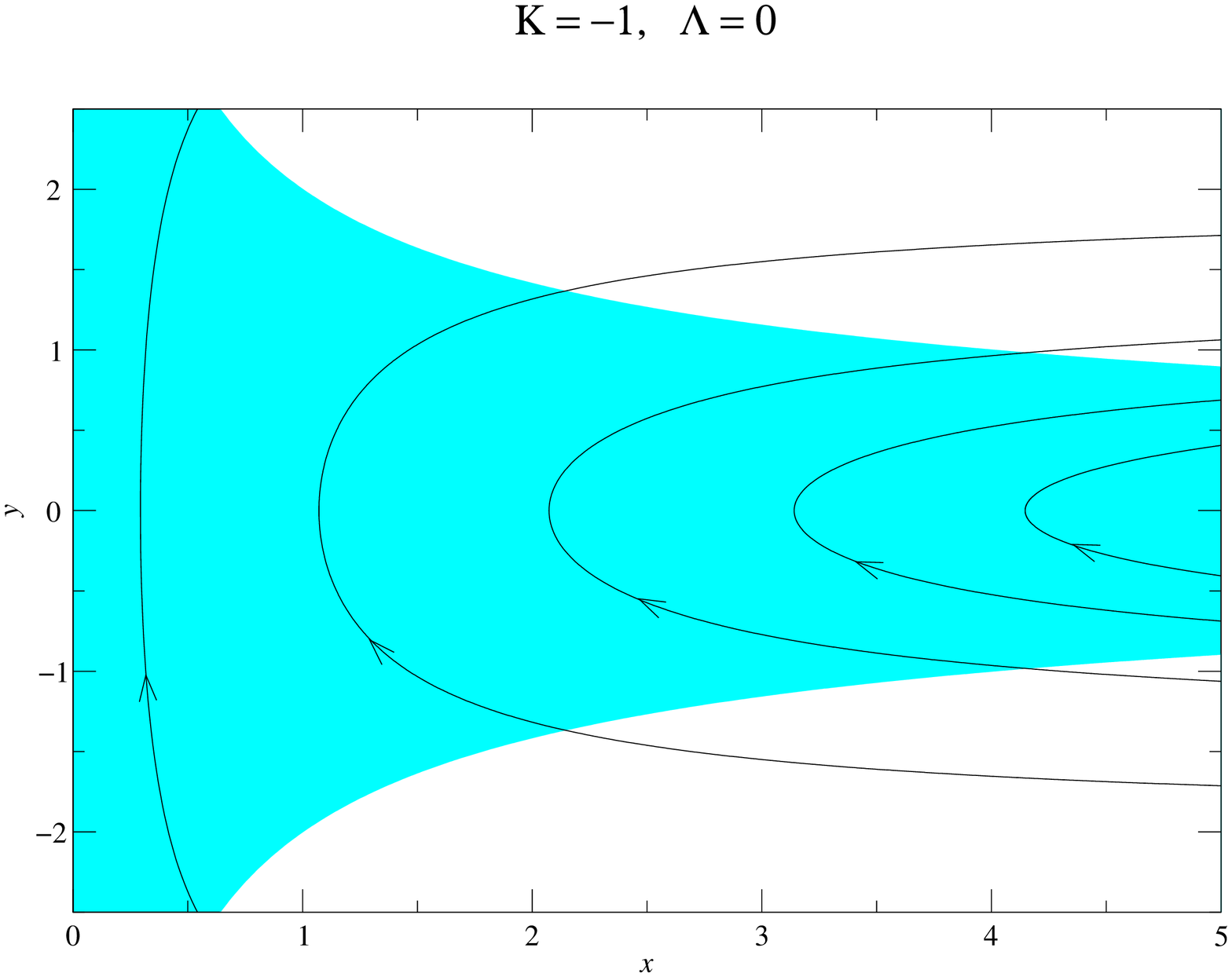}
\includegraphics[angle=-90,width=0.4\textwidth]{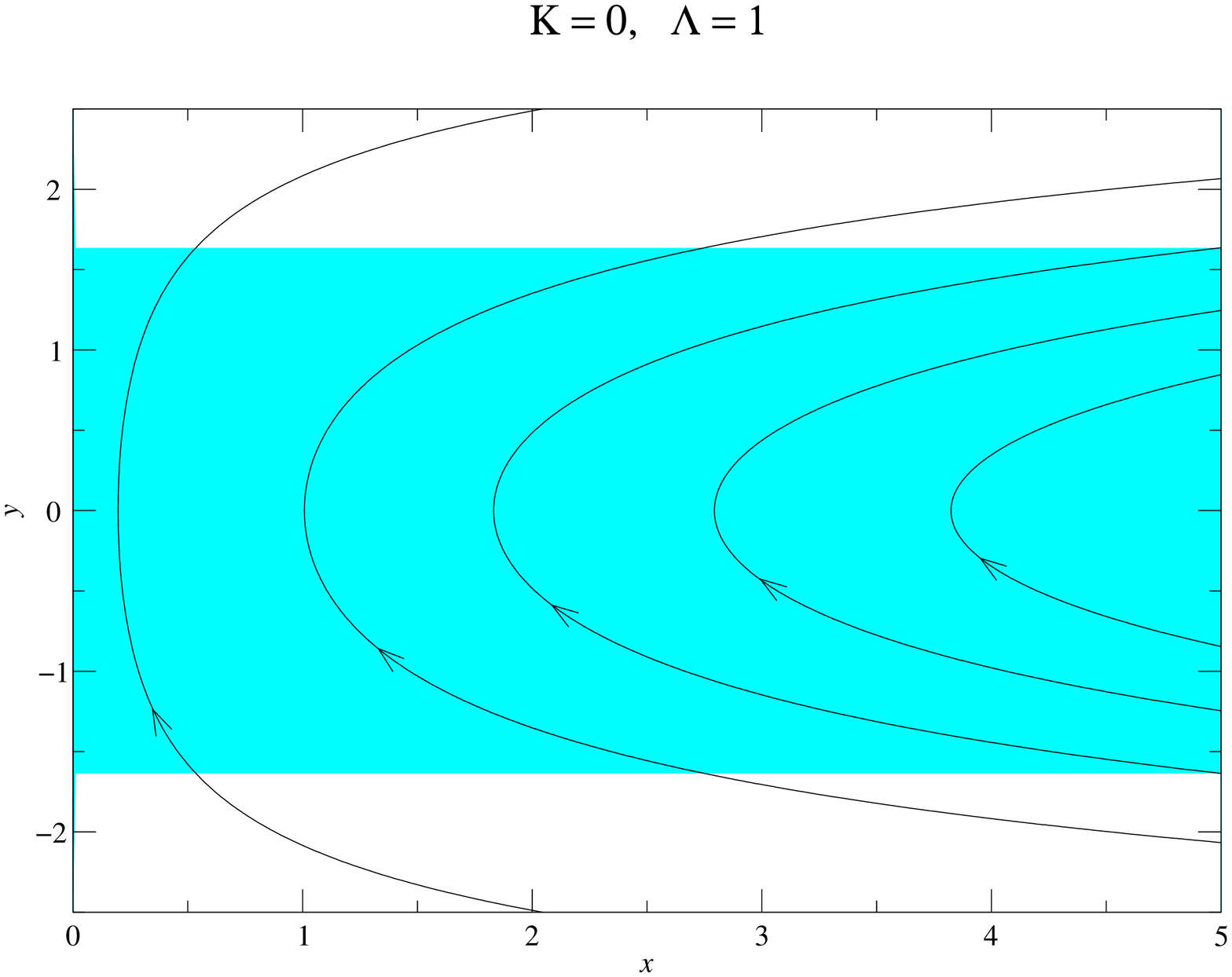}%
\includegraphics[angle=-90,width=0.4\textwidth]{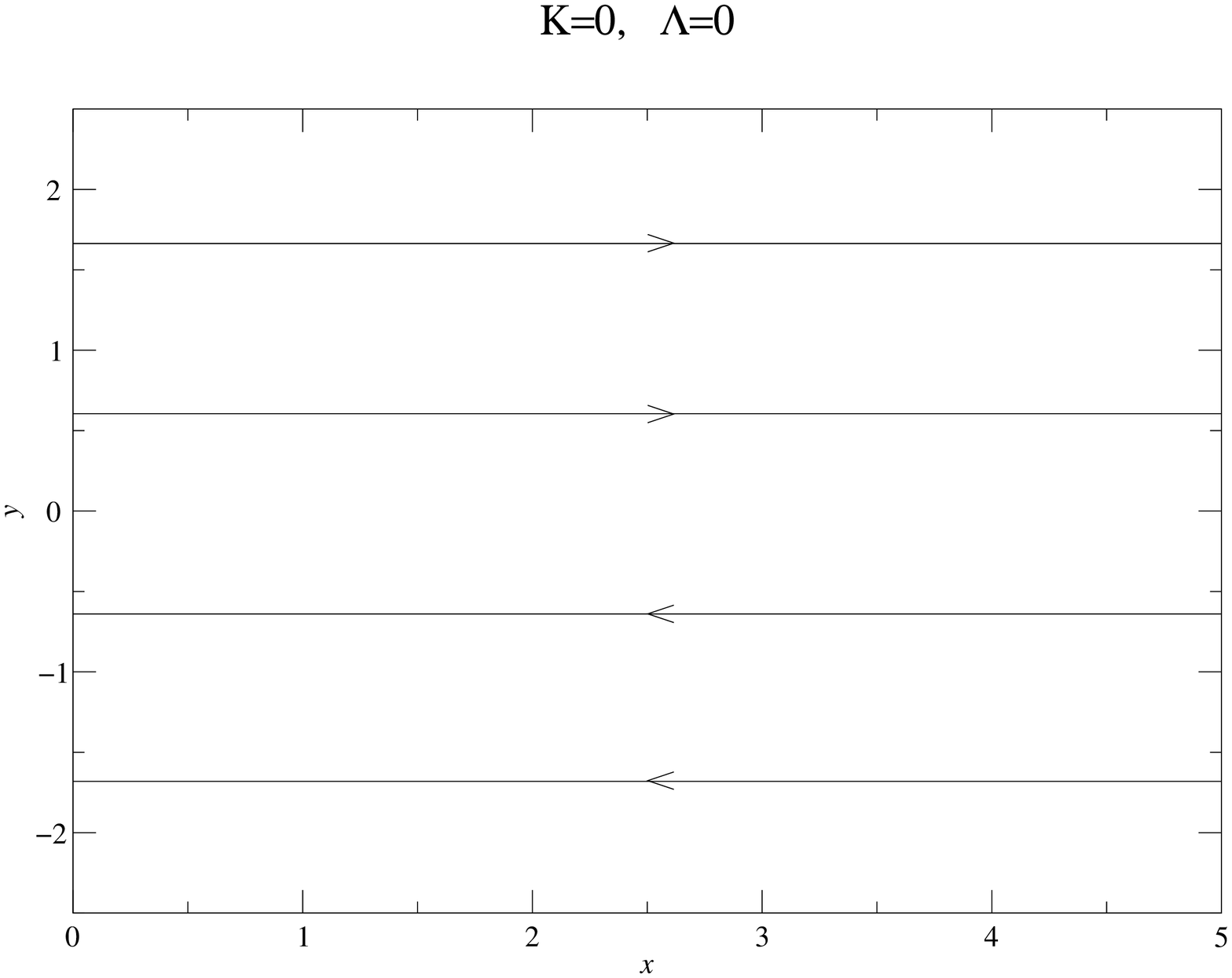}
\includegraphics[angle=-90,width=0.4\textwidth]{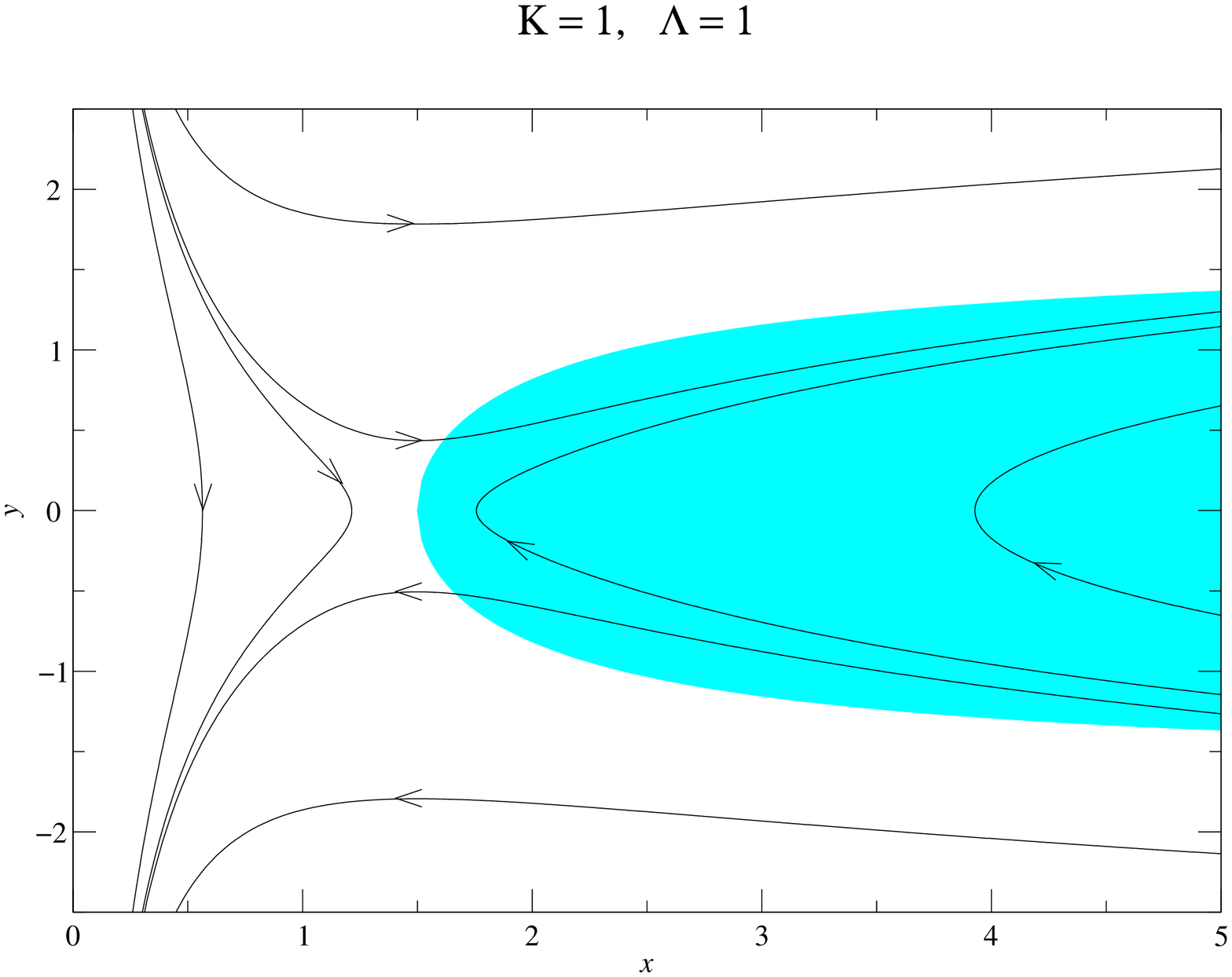}%
\includegraphics[angle=-90,width=0.4\textwidth]{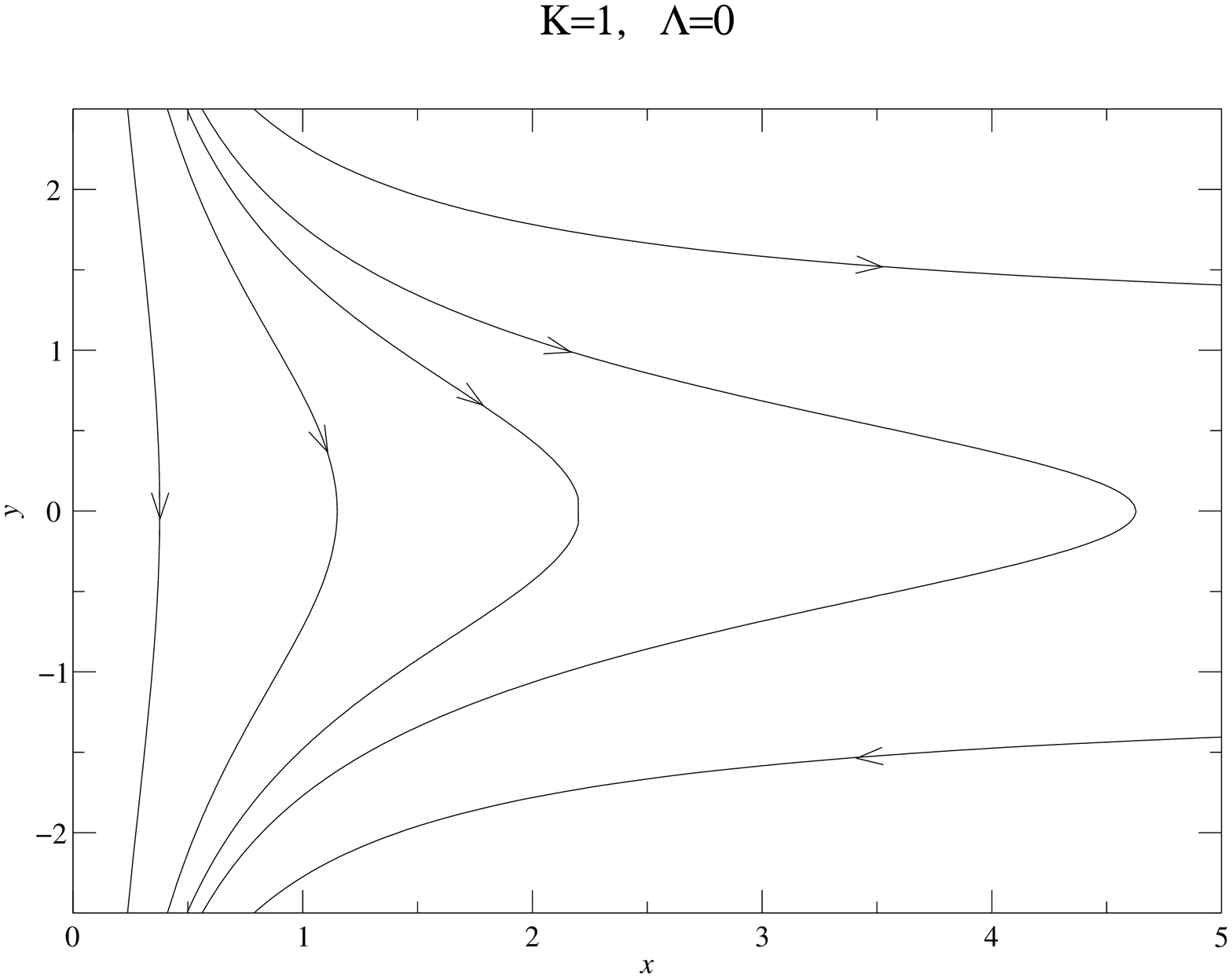}
\caption{The phase portrait for system (\ref{eq:64})--(\ref{eq:65}) for 
$n=-2$. For the cases ($K=-1, \Lambda=1$), ($K=-1, \Lambda=0$), 
($K=0, \Lambda=1$) the qualitative structure of the phase space is the 
same. In the finite domains of phase space there are no critical points. 
Typical trajectories represent the solution starting from singularity-free 
stage $x=\infty$ at $t=0$, then reach the stage $x=x_{0}, y=0$ and goes 
to infinity. In this three cases the trajectories pass through the 
acceleration region (the shaded region). The largest acceleration region 
is for the ($K=-1, \Lambda=1$). These models accelerate for finite interval 
of time, and this acceleration happens to the models without the cosmological 
constant. The closed models for ($K=1, \Lambda=0$) are the typical 
oscillating models. There are no acceleration for ($K=1, \Lambda=0$), 
($K=0, \Lambda=0$). In the case of ($K=1, \Lambda=1$) the saddle appears 
and we have two additional types of evolution. The Lema{\^i}tre model with 
quasi-static stage of evolution and models realized in the aforementioned 
opened and flat models. The acceleration region is in the middle of 
quasi-static phase of evolution of the Lema{\^i}tre.}
\label{fig:1}
\end{figure}

In the above system instead of $a$ in the spirit of the first 
approach we introduce a new variable 
$x \equiv a^2$ and we obtain the system in a simpler form 
\begin{align}
\label{eq:64}
\dot{x} &= y \\
\label{eq:65}
\dot{y} &= - 2\left( K - \frac{2}{3} \Lambda x \right) c^{2}(a(x)) 
\end{align}
where $c^{2}(a(x)) = c_{0}^{2} a^{2n} = c_{0}^{2} x^{n}$. 
The phase portraits of system (\ref{eq:64})-(\ref{eq:65}) 
for $n=-2$ are presented on Fig.~\ref{fig:1}

In the projective coordinates $(z,u)$ the above system takes 
the form 
\begin{align}
\label{eq:66}
\dot{z} &= - zu \\ 
\label{eq:67}
\dot{u} &= - 2\left( K z - \frac{2}{3} \Lambda \right) c_{0}^{2} 
z^{-n} - u^{2} 
\end{align}
This form of the system is useful in analysis of behaviour of trajectories 
at infinity, because $z=0$ corresponds a circle at infinity $x=\infty$ 
which bounds the phase plane. 
The phase portraits of system (\ref{eq:66})-(\ref{eq:67}) 
for $n=-2$ are presented on Fig.~\ref{fig:1}

\begin{figure}
\includegraphics[angle=-90,width=0.4\textwidth]{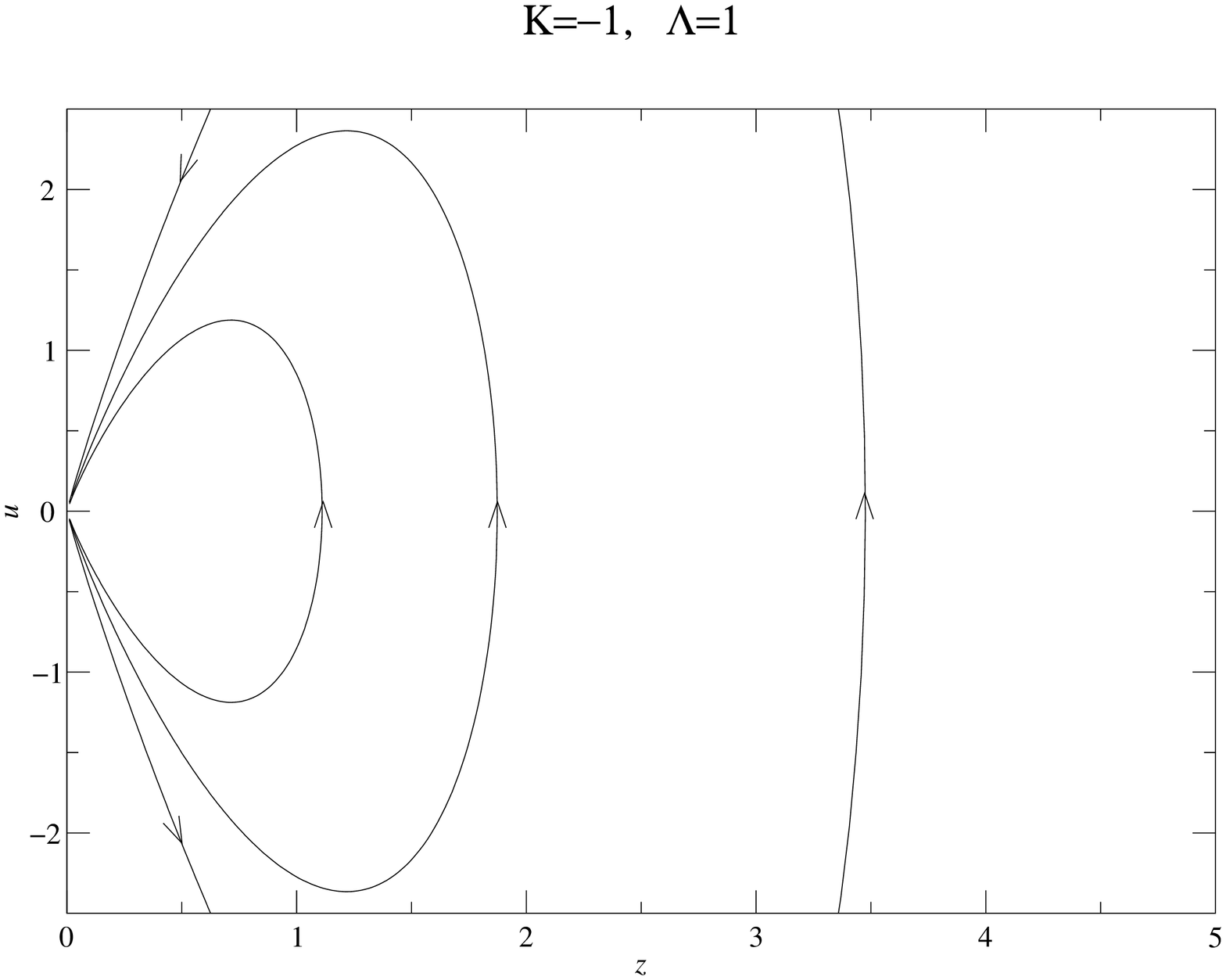}%
\includegraphics[angle=-90,width=0.4\textwidth]{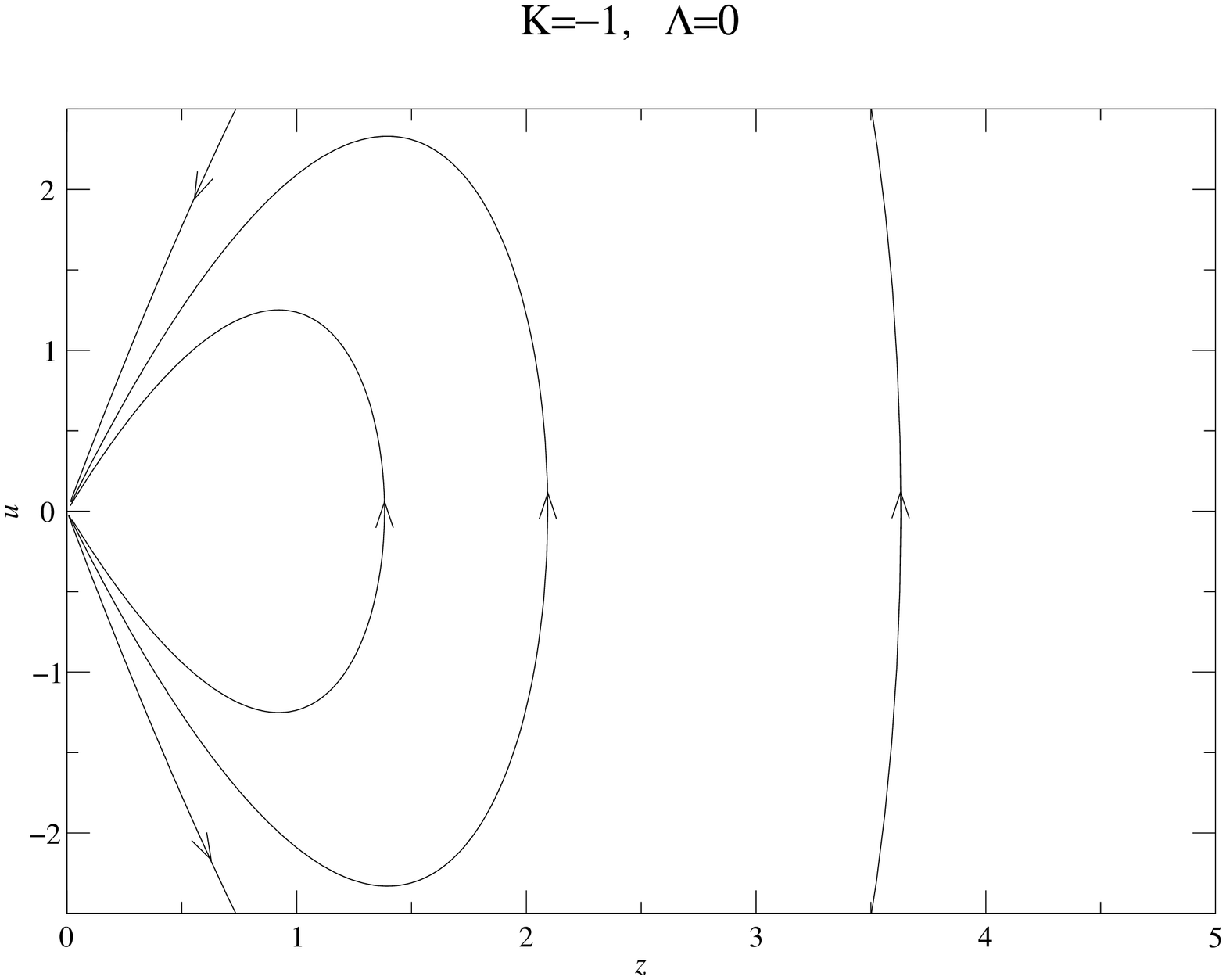}
\includegraphics[angle=-90,width=0.4\textwidth]{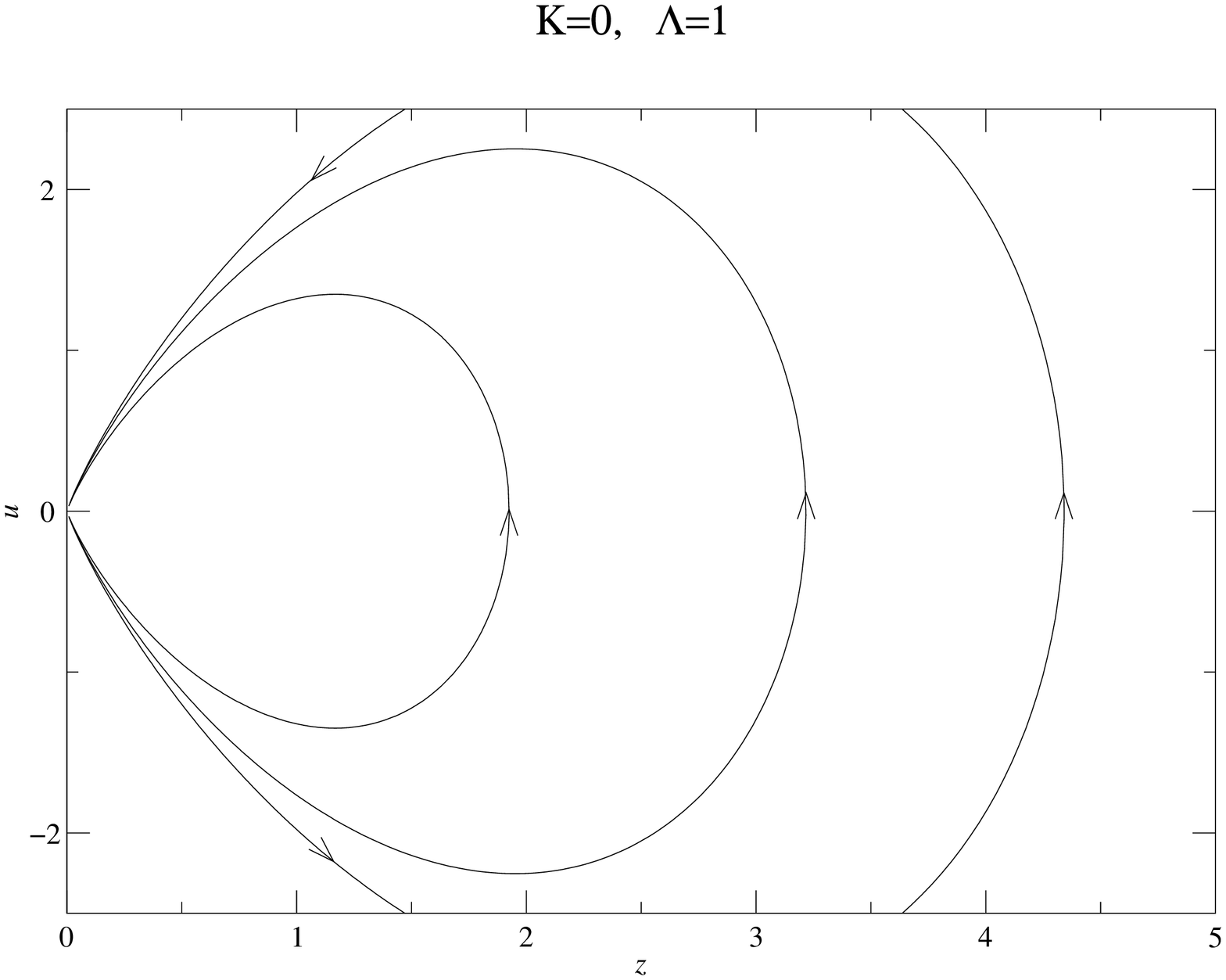}%
\includegraphics[angle=-90,width=0.4\textwidth]{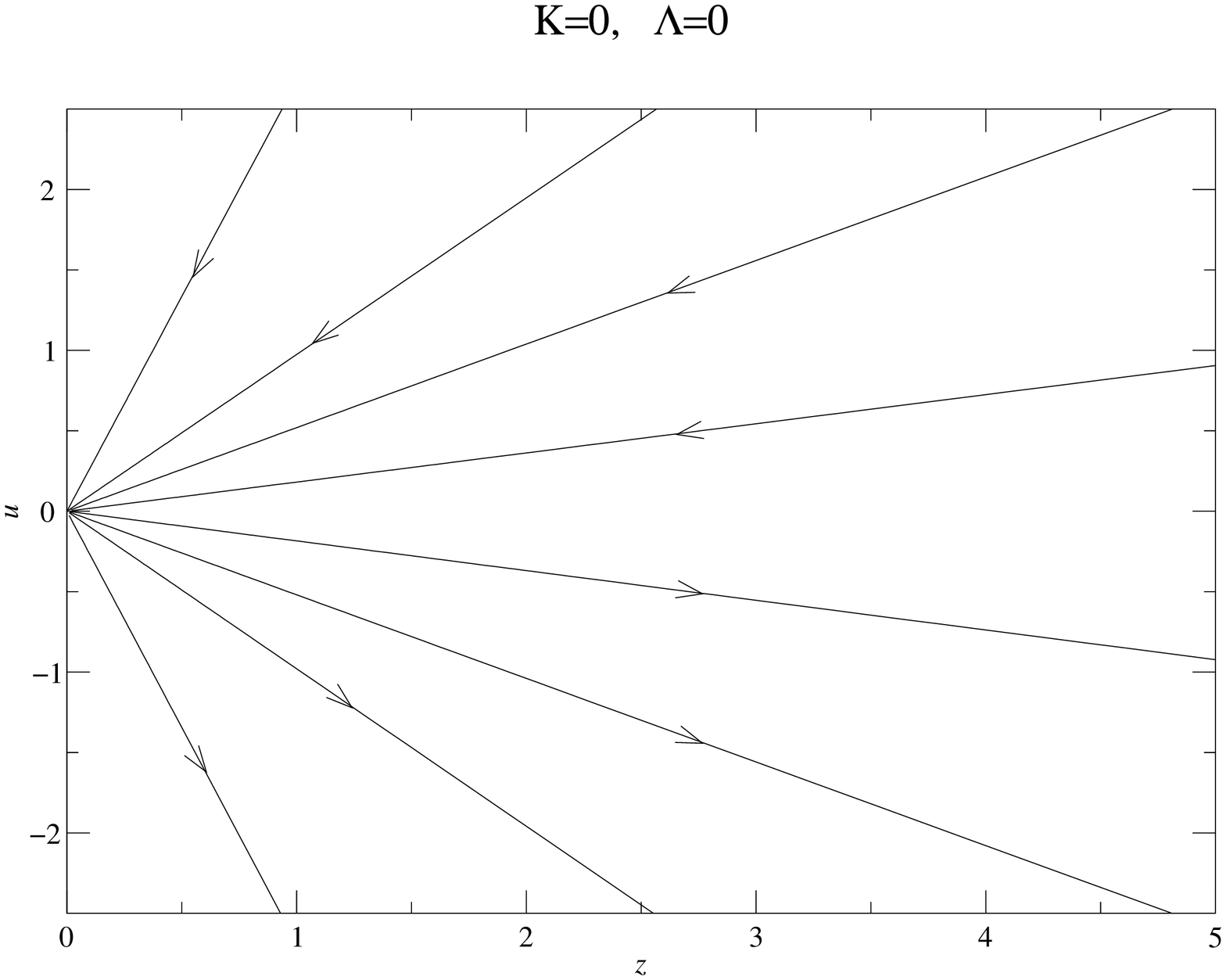}
\includegraphics[angle=-90,width=0.4\textwidth]{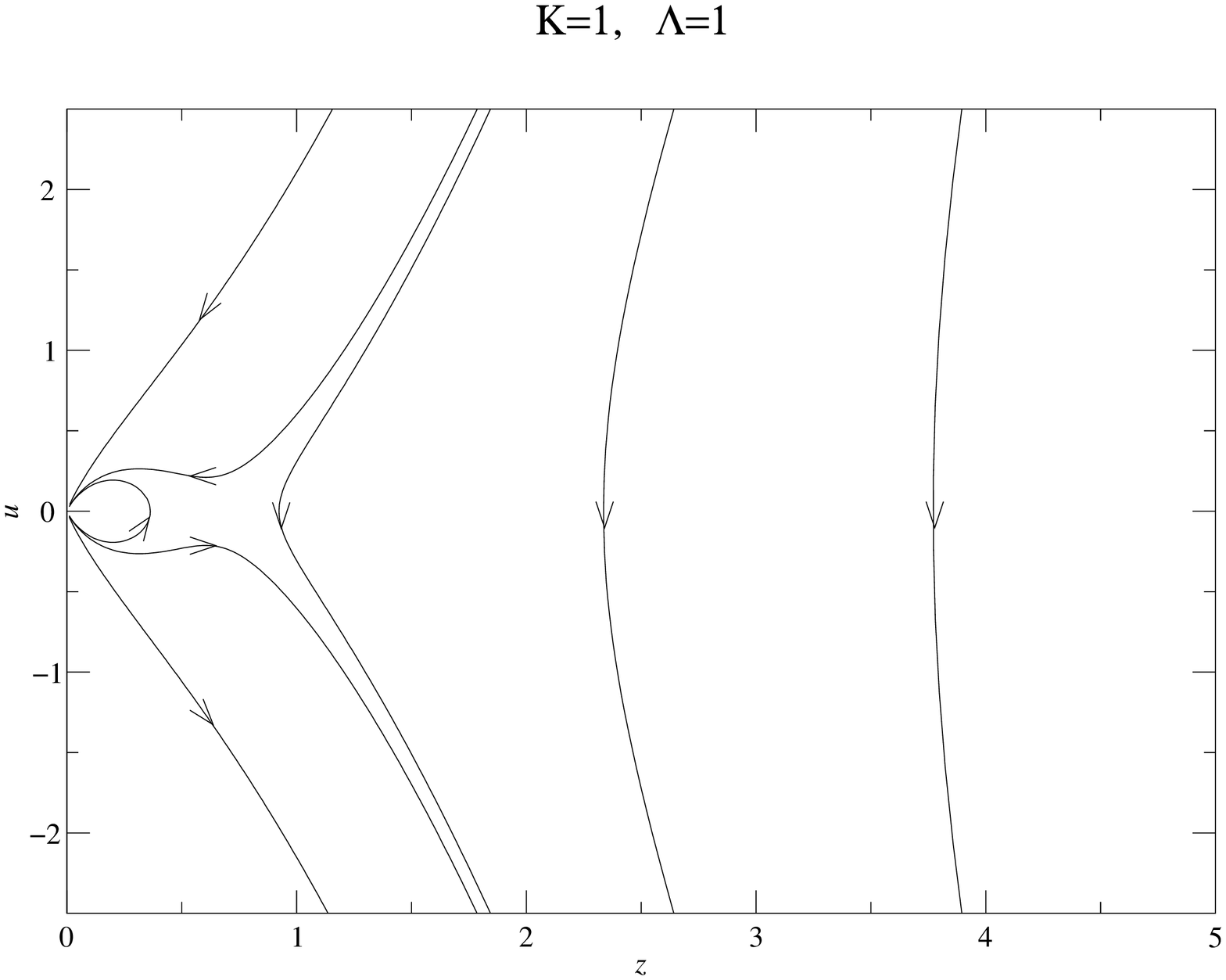}%
\includegraphics[angle=-90,width=0.4\textwidth]{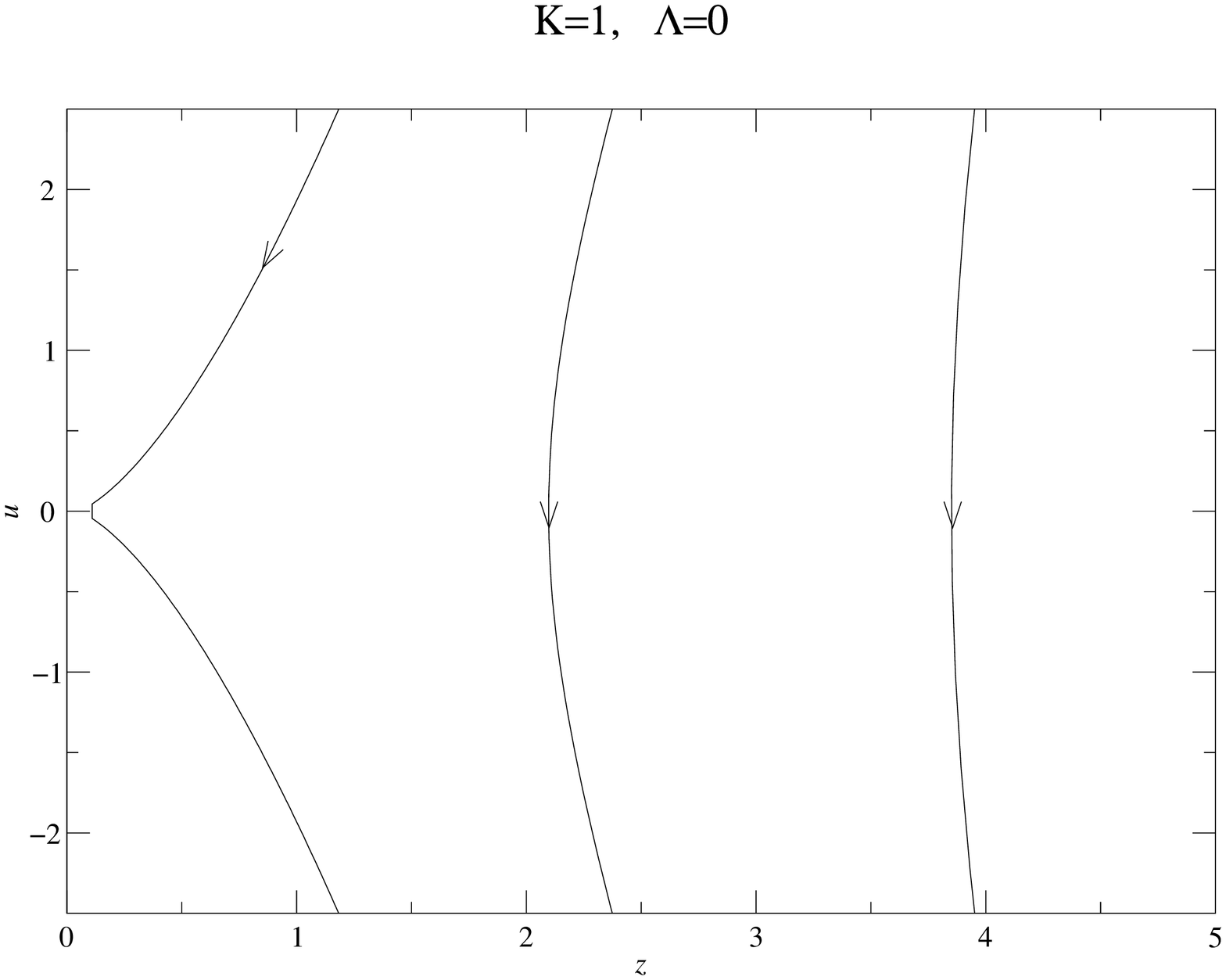}
\caption{The phase portrait for system (\ref{eq:66})--(\ref{eq:67}) for 
$n=-2$ in the projective coordinates $z = 1/x$, $u=y/x$. All critical 
points at infinity corresponds to $z_0 =0$. There are two types of 
critical points $(z_0,u_0) = (0,0)$ or $(z_0,u_0) = (2\Lambda/3K,0)$. 
For all points $\tr A =0$ (where $A$ is a linearization matrix of the 
considered system), i.e., at least one of the eigenvalues is zero. 
If ($\Lambda > 0$, $K=1$) we have a saddle point at $(z_0,0)$. 
The presence of degenerate points at infinity $(0,0)$ that the system 
is structurally unstable in contrast to the models with positive 
cosmological constant and constant velocity of light where 
$(0,\sqrt{4\Lambda/3})$ represents a stable node (the de Sitter stage). 
For $n>0$ there is no critical points at $(0,0)$ and models are 
structurally stable.}
\label{fig:2}
\end{figure}

In this case first integral~(\ref{eq:22}) takes the form 
\begin{equation}
\label{eq:68}
\frac{\dot{x}^2}{2} + 2 \int \left( K - \frac{2}{3} 
\Lambda x \right) c^{2}(x) dx = \bar{C} = \text{const}.
\end{equation}

Let us note that in the special cases of $n=-1$ and $n=-2$ the potential 
takes the particular form $V(x)=2(K\ln x - \frac{2}{3} \Lambda x)$, 
$V(x) = 2(-\frac{K}{x} - \frac{2}{3} \Lambda \ln x)$.

It is clear that first integral (\ref{eq:68}) is in fact 
the integral of energy because system (\ref{eq:64})--(\ref{eq:65}) 
is a Hamiltonian dynamical system with the Hamiltonian 
\[
{\cal H}(p,x) = \frac{p^{2}}{2} + 2 \left( \frac{K x^{n+1}}{n+1} 
- \frac{2}{3} \Lambda \frac{x^{n+2}}{n+2} \right) \equiv \bar{C} = 
\text{const} > 0.
\]
Now the integral of energy can be used in the classification 
of all possible evolution modulo their quantitative (i.e., in accuracy 
to differential type) properties. 

In the qualitative classification, for any case of $\gamma(a)$, first 
integral~(\ref{eq:22}) may be useful as in the method of the 
classification previously used in \cite{Szydlowski84}. It assumes the form 
\begin{equation}
\label{eq:69}
\frac{p^{2}}{2} + V(a(x)) = V(a_{0})
\end{equation}
where
\[
V(a) = \int^{a(x)} \frac{a^{2}}{\alpha a+1} \left[ \frac{K(1+3\gamma(a))}{2a} 
- \Lambda a \left( 1 + \frac{1}{3(\alpha a+1)} \right) 
\right] c^{2}(a) da ,
\]
$V(a_{0}) = \text{const} >0$ and $x = \frac{2\sqrt{2}}{\alpha^{2}} 
t^{2}(t-1)$, $t=\sqrt{\alpha a+1}$ and $x=a^{2}$ for $\alpha =0$. 

From equation~(\ref{eq:69}), after imposing the condition $p = 0$, 
we can calculate $\Lambda$ from the expression on function $\Lambda(a)$ which 
constitutes a boundary of a domain of configuration space admissible 
for motion 
\[
p^{2} \ge 0 \quad \Leftrightarrow \quad 
V(a) - V(a_{0}) < 0 
\]
and 
\begin{equation}
\label{eq:70}
\Lambda(a) \ge \frac{\int^{a} \frac{a(\alpha a +2)}{2(\alpha a +1)^{2}} 
K c^{2}(a) da - V(a_{0})}{\int^{a} \frac{(3\alpha a + 4) 
a^{2}}{3(\alpha a+1)^{2}} c^{2}(a) da}.
\end{equation} 

By consideration the boundary $\partial D \colon \{ (a,\Lambda) \colon 
\Lambda(a) = 0 \}$ and construction of the levels $\Lambda = 
\text{const}$ we obtain the qualitative classification of 
all possible trajectories in the space $(\Lambda,a)$ or 
$(\Lambda,x)$.

\subsection{How to interpret the acceleration of scale 
factors and the absence of particle horizon}

A great advantage of the phase-space dynamical description is the ability 
to discuss the distribution of models with given properties. In the other 
words one can imagine an ansamble of models starting from different 
initial conditions and ask how a given property is 
distributed in the ansamble. Now we formulate sufficient conditions 
for solving the flatness and horizon problems in terms of phase-space 
relations. Let us recall that the flatness problem is solved whenever 
the scale factor's acceleration is positive 
\[
\ddot{a}(t) > 0.
\]
This condition is fulfilled in the subspace ${\cal D}_{\text{flat}}$ 
of the phase space 
\begin{equation}
\label{eq:71}
{\cal D}_{\text{accel}} = \left\{ (a,\dot{a}) \colon 
- \frac{\alpha a + 2}{2a(\alpha a + 1)} p^{2} - 
\left( \frac{K(\alpha a +2)}{2a(\alpha a + 1)} 
- \frac{3 \alpha a + 4}{6(\alpha a + 1)} 
\Lambda a \right) c^{2}(a) 
> 0 \right\}.
\end{equation}
It means that trajectories representing the histories of VSL 
universes undergo an accelerated expansion while staying 
in region ${\cal D}_{\text{flat}}$. One can restate relation 
(\ref{eq:71}) using the Hamiltonian constraint  
$p^{2} = 2(V(a_{0}) - V(a))$. It is easy to see that the respective 
condition expressed purely in terms of configuration space, 
reads
\begin{equation}
\label{eq:72}
{\cal D}_{\text{accel}} = \left\{ a \colon 
- \frac{\alpha a +2}{a(\alpha a +1)} (V(a_{0}) - V(a))
- \left( \frac{K(\alpha a +2)}{2a(\alpha a + 1)} 
- \frac{3 \alpha a + 4}{6(\alpha a + 1)} 
\Lambda a \right) c^{2}(a) 
> 0 \right\}.
\end{equation}
Analogous criteria of acceleration if dynamics is covered by equations 
(\ref{eq:54})--(\ref{eq:55}) are given by (\ref{eq:57}) in phase space and 
(\ref{eq:59}) in the configuration space. 

Another interesting question concerns the horizon problems. It is 
easy to prove the following criterion of avoiding the horizon problem.
\begin{theorem}
The FRW cosmological model does not have an event horizon near 
the singularity if $\dot{a}(t)c^{-1}(t)$ tends to a constant while 
$a(t)$ tends to zero.
\end{theorem}
{\em Proof.\/} When all events whose coordinates at past time are 
located beyond some distance $d_H$ they can never communicate with the 
observer at coordinate $r=0$ in the Robertson-Walker metric we can 
define the distance $d_H$ as past event horizon distance. 
It is given by 
\[
d_{H}(t) = a(t) \int_{0}^{t_{0}} \frac{dt' c(t')}{a(t')} = a(t)I.
\]
Whenever $I$ diverges as $t \to 0$ there is no past event horizon 
in the space time geometry. On the other hand, when $I$ converges 
the space time exhibits a past horizon
\[
\int_{t}^{t_{0}} \frac{dt c(t)}{a(t)} = 
\int_{t}^{t_{0}} \frac{a^n da}{a \frac{da}{dt}} =
\int_{t}^{t_{0}} \frac{1}{c^{-1}\dot{a}}\frac{da}{a}. 
\]
Let $c^{-1}(t)\dot{a} < A$ then 
\[
I \ge \frac{1}{A} \int_{0}^{a_{0}} \frac{da}{a} 
= \frac{1}{A} ( \ln a_{0} + \infty).
\]
On the other hand when $\dot{a} < \bar{A}$ is bounded then 
\[
I = \int_{t}^{t_0} \frac{dt c(t)}{a(t)} 
= \int_{t}^{t_0} \frac{a^{n} da}{a \frac{da}{dt}} 
= \int_{t}^{t_0} a^{n-1} \frac{da}{\dot{a}} 
\]
and
$I \ge \frac{1}{\bar{A}} \int_{0}^{a_0} a^{n-1} da$
or 
\[
\bar{A} \int_{t}^{t_0} \frac{dt c(t)}{\dot{a}(t)} \ge \frac{a^{n}}{n}.
\]
Therefore $I$ diverges as $a \to 0$ if $n<0$ and $\dot{a}< \bar{A}$ $\Box$.

The above criterion can be reformulated in the language of the phase 
space in the form 
\[
a \to 0 \text{ and } c^{-1}(t) \dot{a}(t) \to \text{const}, 
\quad \text{i.e. } a^{-2n} \left( \frac{da}{dt} \right)^2 \to 
(\text{const})^{2}.
\]
For example for the radiation case $a=\sqrt{x}$ and then the past horizon 
is eliminated if only $\dot{x}^{2} x^{-2n-1} \to \text{const}$ as 
$x \to 0$. After substituting first integral (\ref{eq:68}) we obtain 
that there is such $n$ that the horizon disappears if only if $n < -1$. 
Let us note that the above proof is based on the Hamiltonian constraint 
and is independent of any specific assumption about an equation of state or 
$a(t)$ near the singuarity. If we assume the power-law behaviour of $a(t)$ 
then Barrow's result can be simply achieved \cite{Barrow02}, namely 
$a(t) \propto t^{2/3(\gamma - 1)}$ and if $2 \ge 2n + 3(\gamma +1)$.

\subsection{The potential function for mixture of dust and radiation}

The boundary of the domain in the configuration space ($x$-space or 
$a$-space) admissible for motion is determined by expression for 
\[
V(a(x)) = Kc_{0}^{2} \int^{a(x)} 
\frac{a^{1+2n}(\alpha a+2)}{(\alpha a+1)^2} 
da - \Lambda c_{0}^{2} 
\int^{a(x)} \frac{(3\alpha a + 4)a^{3+2n}}{3(\alpha a+1)^{2}} da 
= V(a_{0}) > 0.
\]
The equation $V(a(x)) - V(a_{0}(x_{0})) =0$ can be represented 
in the space $(\Lambda,a)$ or more correctly in the space 
$(\Lambda,x)$ as a boundary curve for the classification. Instead of the 
inverse function $a(x)$ (it is difficult to give it in a simple form), 
we take the function $x(a)$ as
\[
V(a_{0}) - K c_{0}^{2} \int^{a} \frac{a^{1+2a}}{2(\alpha a+1)}da 
- K c_{0}^{2} \int^{a} \frac{a^{1+2n}}{2(\alpha a+1)^{2}} da 
+ \Lambda c_{0}^{2} \left[ \int^{a} \frac{a^{3+2n}}{3(\alpha a+1)} da 
+ \frac{1}{3} \int^{a} \frac{a^{3+2n}}{(\alpha a+1)^{2}} da \right] = 0 
\]
where the physical domain is $V(a_{0}) - V(a) >0$, i.e. 
\[
\Lambda \ge \frac{-V(a_{0}) + \frac{1}{2}K c_{0}^{2} 
\int^{a} \frac{a^{1+2n}}{\alpha a+1} da + \frac{1}{2}Kc_{0}^{2} 
\int^{a} \frac{a^{1+2n}}{(\alpha a+1)^{2}}da}{c_{0}^{2} 
\left[ \int^{a} \frac{a^{3+2n}}{3(\alpha a+1)} da 
+ \frac{1}{3} \int^{a} \frac{a^{3+2n}}{(\alpha a+1)^{2}} da \right]}.
\]
where 
\begin{eqnarray*}
\int \frac{da}{a^{m}(\alpha a+1)} &=& 
\sum_{k=1}^{m-1} \frac{(-1)^{k} \alpha^{k-1}}{(m-k)! a^{m-k}} 
+ (-1)^{m} \alpha^{m-1} \ln \frac{\alpha a+1}{a} \\
\int \frac{da}{a^{l}(\alpha a+1)^{2}} &=&
- \frac{1}{\alpha} \left[ \frac{1}{a^{l} (\alpha a+1)} 
+ l \left( \sum_{k=1}^{l} \frac{(-1)^{k} \alpha^{k-1}}{(l+1-k)! a^{l+1-k}} 
+ (-1)^{l+1} \alpha^{l} \ln \frac{\alpha a+1}{a} \right) \right].
\end{eqnarray*}

\begin{figure}
\includegraphics[width=0.8\textwidth]{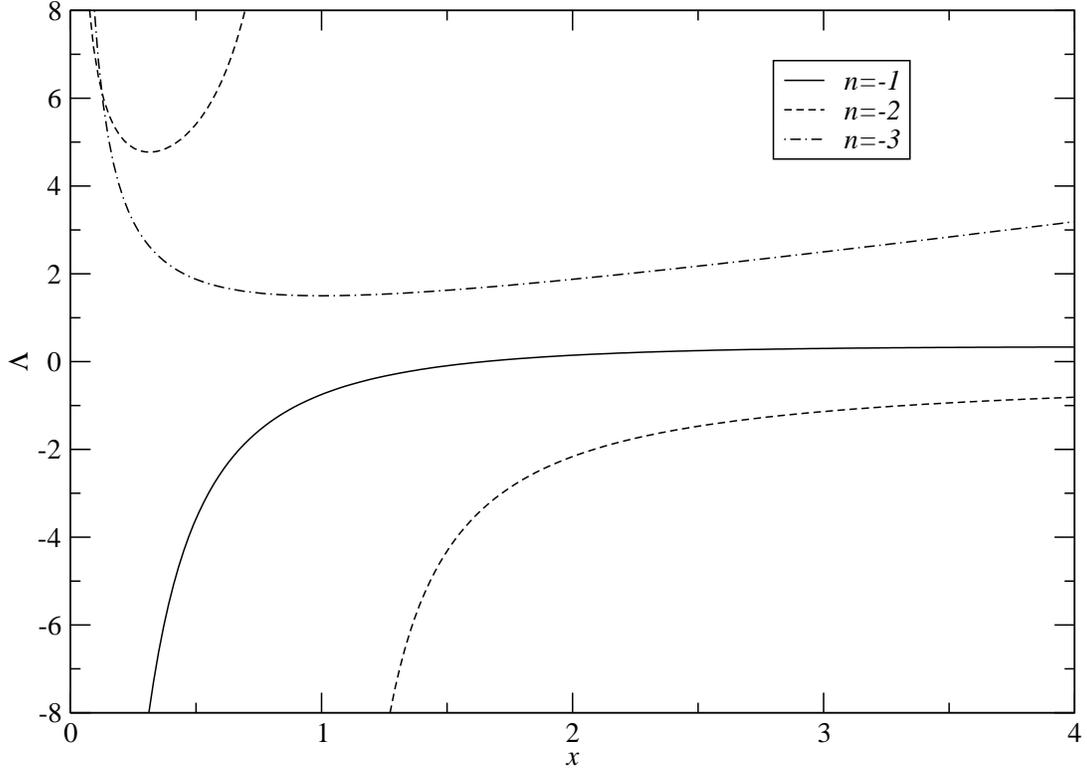}
\caption{The relationship between $\Lambda$ and $x$ for $K=1$. 
The qualitative evolution of the models is represented by levels of 
$\Lambda = \text{const}$. The domain under the characteristic curves 
$\Lambda(a)$ is classically forbidden. For $n=-1$ and $\Lambda > 0$ 
all models start from finite size of scale factor and expand. 
For $n=-1$ and $\Lambda>0$ all models start from the singularity 
and expand. For $n=-2$ and $\Lambda>0$ all models start from finite 
scale factor and expand. For $n=-2$ and $\Lambda$ there are oscillating 
universes and with a singularity and other universes expanding from 
finite scale factor. For $n=-3$ all models are oscillating and with 
a singularity.}
\label{fig:3}
\end{figure}

\begin{figure}
\includegraphics[width=0.8\textwidth]{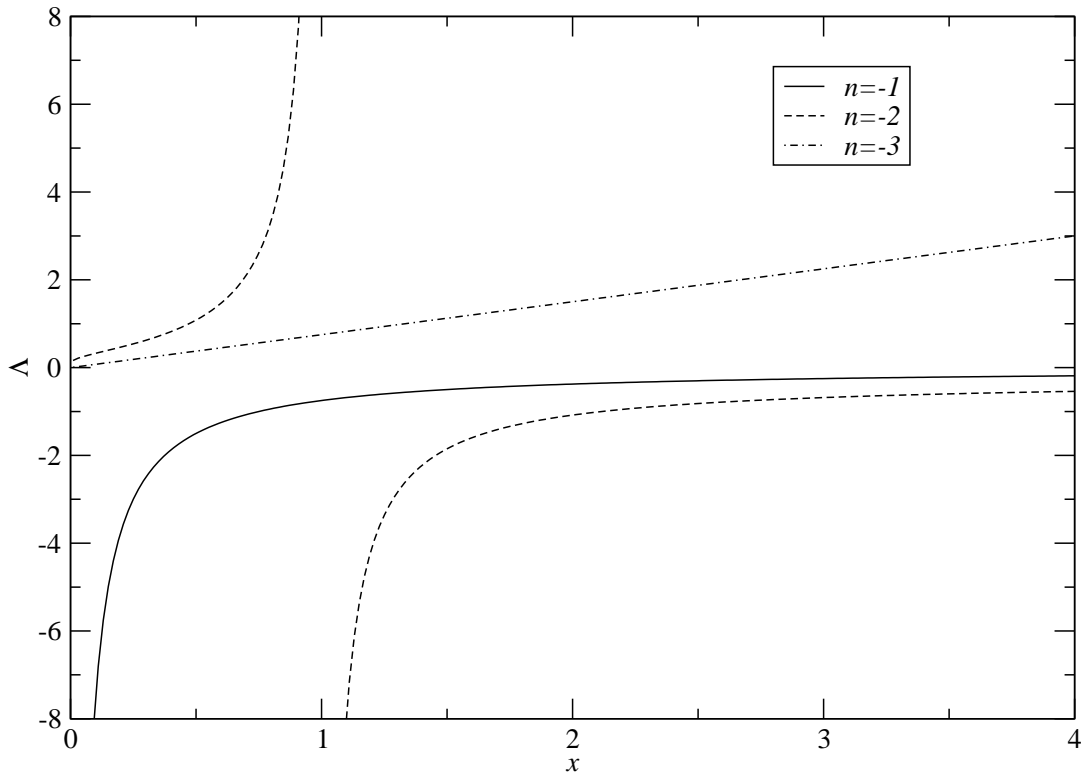}
\caption{The relationship between $\Lambda$ and $x$ for $K=0$. 
The qualitative evolution of the models is represented by levels of 
$\Lambda = \text{const}$. The domain under the characteristic curves 
$\Lambda(a)$ is classically forbidden. For $n=-1$ and $\Lambda < 0$ 
all models oscillate. For $n=-1$ and $\Lambda > 0$ models 
evolve from a singularity to infinity. For $n=-2$ and $\Lambda < 0$ 
models oscillate. For $n=-2$ and $\Lambda > 0$ models also 
oscillate but without a singularity. For $n=-3$ and $\Lambda < 0$ 
models expand from a singularity to infinity. For $n=-3$ and $\Lambda > 0$ 
all models oscillate. 
}
\label{fig:4}
\end{figure}

\begin{figure}
\includegraphics[width=0.8\textwidth]{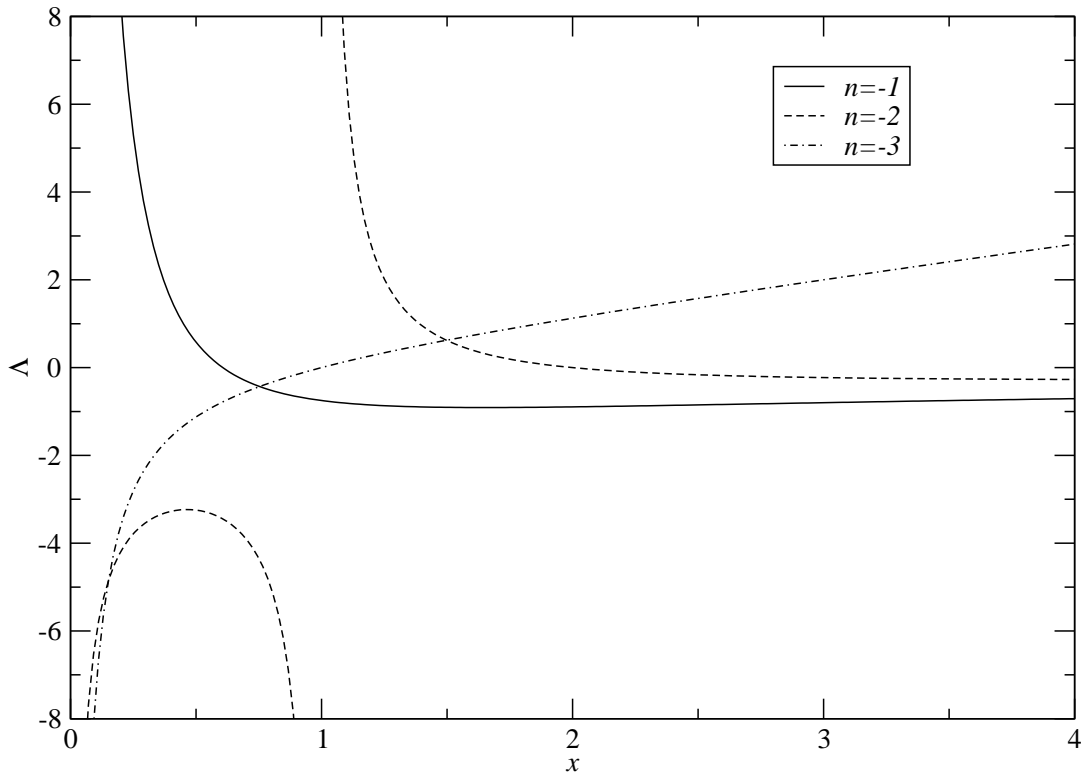}
\caption{The relationship between $\Lambda$ and $x$ for $K=-1$. 
The qualitative evolution of the models is represented by levels of 
$\Lambda = \text{const}$. The domain under the characteristic curves 
$\Lambda(a)$ is classically forbidden. For $n=-1$ and $\Lambda < 0$ 
there is no solution. For $n=-1$ and $\Lambda > 0$ models start from 
finite size and expand. For $n=-2$ there is discontinuity at $x=1$ 
and for $\Lambda < 0$ some models oscillate from a singularity and 
other models oscillate without a singularity. For $n=-2$ and $\Lambda > 0$ 
models start from finite size of scale factor to $x = \infty$. 
For $n=-3$ all models oscillate with a singularity.
}
\label{fig:5}
\end{figure}

On the figures~\ref{fig:3}--\ref{fig:5}, for the simplicity of presentation 
without loss of generality, it is assumed $c_{0}=1$, $V(a_{0})=1$ and $K$, 
$\alpha$ are parameters and $n$ is chosen $-1$, $-2$, $-3$.

\section{Tunnelling in $n$ decaying cosmologies}

In the classical VSL cosmology a particle trajectory is determined 
through a knowledge of both a position $x$ and a canonically conjugate 
momentum $p_x$. In the quantum VSL cosmology the notion of trajectories 
loses its classical meaning due to the uncertainty relation ($x$ and 
$p_x$ are replaced by non-commuting operators). The Wheeler-De Witt 
equation 
\begin{equation}
\label{eq:91}
\left[ \frac{\partial^2 \phantom{x}}{\partial x^2} - V(x) \right] 
\psi(a) = 0
\end{equation} 
is identical to the one-dimensional time-independent Schr\"odinger 
equation for a one half unit particle of energy $E$ subject to 
the potential $V(x)$ ($\psi(a)$ is known as the wave function of 
the universe).

The `particle-universe' can quantum-mechanically tunnel through 
the potential barrier. Let us consider the case when the region 
beneath the barrier $0 < a < a_0$, is classically forbidden 
whereas the region $a \ge a_0$ is classically allowed. 

We can adopt a simple method of calculating an amplitude for 
the quantum creation of the VSL FRW universe from a zero size 
to $a = a_0 = \sqrt{\frac{3(n+2)}{2(n+1)\Lambda}}$ 
\begin{equation}
\label{eq:92}
| \text{VSLFRW}(a_0) | \text{nothing} \rangle |^2 = P 
\cong \exp\left[- \frac{2}{\hbar} \int_{0}^{x_0} \sqrt{2(E-V)} dx \right].
\end{equation}
The above formula (called Gamov's formula) gives us tunnelling 
probability because the quantized VSL FRW universe is mathematically 
equivalent to a one-dimensional particle of unit mass. 

As an example, let us get back to the previously considered case of 
the compact vacuum VSL model. The Hamiltonian for this case can be 
obtained if we put $E=0$ into $\mathcal{H}(p,x)$. 

The region of barrier $0<x<x_0$ is classically forbidden for the zero 
energy particle. Therefore one can find the probability that particle 
at $x=0$ can tunnel to $x=x_0 \colon V(x_0)=0$; 
$x_0 = \frac{3(n+2)}{2(n+1)\Lambda}$. 

After rescaling the variable $x \mapsto x/x_0 = \kappa$ we obtain for 
(\ref{eq:92}) 
\begin{equation}
\label{eq:93}
P \cong \exp\left[ -\frac{4}{\hbar}\left( \frac{3(n+2)K}{2(n+1)\Lambda} 
\right)^{(n+3)/2} \right] \int_{0}^{1} \kappa^{n/2} 
\sqrt{\kappa (1 - \kappa)} d\kappa
\end{equation}
where we assume $K=1$, $\Lambda \ge 0$ and $-1 <n < 0$; the potential 
$V(x)= x^{n+1}(1+x)$ has two extrema and two zeroes. Relation 
(\ref{eq:93}) can be rewritten to the form 
\begin{equation}
\label{eq:94}
P \cong \exp\left[ - \frac{2}{\hbar} F(n) \right]
\end{equation}
where 
\begin{equation}
\label{eq:95}
F(n) = \left[ \frac{3(n+2)}{2(n+1)\Lambda} 
\right]^{\frac{n+3}{2}} \frac{\sqrt{\pi}}{2} 
\frac{\Gamma \left( \frac{n+3}{2} \right)}{\Gamma \left( 
3 + \frac{n}{2} \right)}.
\end{equation}

It is the most probable that the closed and vacuum VSL FRW model 
with $-1 < n < 0$ is born when we have maximum permissible energy 
density or least size $a_0$. It occurs that the creation of 
the universe with constant $c$ ($n=0$) is the most probable when 
classical spacetime emerges via the quantum tunnelling process 
whereas $c(a)$ is a decreasing function during the evolution of 
the universe.

\section{Conclusions}

Let us assume that one takes the idea of the varying speed of light 
seriously as a physical effect that might have happened in the very 
early universe and today is confined to a very narrow range admissible 
by inaccuracy of existing bounds on variability of $c$. One of the 
problems arising then is to see how this 
modification of physics would change the evolution of standard 
Friedmann-Robertson-Walker cosmological models. So far only specific 
qualitative results are known concerning the solution of flatness 
and horizon problems in VSL models. In the present work we attempted 
to extend this qualitative discussion in the sense that by constructing 
phase space portraits of VSL cosmological models we were able to obtain 
a global view of their dynamics. In order to achieve this we have used 
a power law Ansatz for function $c(t)$ and investigated classical Einstein 
equations with $c$ allowed to be a function of time. 

The two procedure of reduction of dynamics are proposed. In the first 
case we reduced the dynamics of VSL models to a two-dimensional Hamiltonian 
dynamical system with a quadratic kinetic energy form and a potential 
function depending on a generalized scale factor. In the second one 
we reparametrized the time variable but the scale factor remains 
a state variable. In both cases the shape of the 
potential and the existence of the energy integral was used to classify 
possible evolutions of VSL models. These possibilities comprise models 
evolving from the singularity to infinity, oscillatory behaviour 
between initial and final singularity, Einstein-de Sitter type models 
evolving from the singularity to the static world, Lema{\^i}tre-Eddington 
type models evolving from the static Einstein solution to infinity, 
models expanding to infinity from the finite size and finally models 
starting and ending with finite scale factors. 

We have dealed with the full global dynamics of VSL models. From 
the theoretical point of view it is important how large the class 
of models without horizon or with solved cosmological puzzles is. 
We call this class of models generic if its inset in the open phase 
is open or non-zero measure. Such a point of view is justified by 
the fact that if the solution of a cosmological puzzle is an 
attribute of a trajectory with a given initial conditions, it should 
also be an attribute of another trajectory which starts with 
neighbouring initial conditions. 

We have shown that the assumed time dependence of the speed of light leads 
to a uniform evolution pattern of VSL models on the phase space. The 
criteria for solving the flatness and horizon problems were formulated 
in terms of the phase space. It is an advantage of the phase space 
approach that one can trace the patterns of evolution for all possible 
initial conditions. We have depicted, on respective phase portraits, 
the regions where the flatness problem is solved. The models where 
the region of initial conditions leading to flatness and horizon problem 
avoidance is large play a distinguished role. From this perspective 
open ($K=-1$) models with positive cosmological constant $\Lambda>0$ 
are preferred in the class of VSL FRW models filled with radiation. 

The formalism presented in this paper can be easily extended to the 
case where the matter content of the model is a mixture of different 
types of matter and to the case of models with shear (e.g. Bianchi I 
or V).

This formalism can be also treated as a starting point of an application 
of quantum cosmology to the description of early stages of evolution 
of the Universe \cite{Vilenkin82,Vilenkin83}. The tunnelling rate, with 
an exact prefactor can be calculated to the first order on $\hbar$ 
for the closed VSL FRW model with a decaying variable velocity of light 
term $c(a)$ The tunnelling probability $P$ can be calculated in the WKB 
approximation given in the $V \gg E$ limit by (\ref{eq:92}). 
We consider the closed vacuum VSL FRW models for which potential is 
qualitatively classical. This implies that $-1<n$. In the interval 
$-1< n \le 0$ the probability of tunnelling increases as $F(n)$ 
monotonically decreases with increasing $n$. It is shown that 
the highest tunnelling rate occurs for $n=0$; it corresponds to the 
standard FRW model. 

In our work we showed the effectiveness of dynamical system methods 
in the investigation of VSL FRW models, namely

In the distinguished class of open models with the cosmological constant 
the acceleration has `transitional' character, i.e., there 
is a finite time when trajectories are in the acceleration region, and 
the measure of this region normalized to the area of phase plane is 
finite, even in the case of $\Lambda=0$. 

We can argue that the considered VSL models are structurally unstable 
(Fig.~\ref{fig:2}) because of the presence of degenerate critical points 
at infinity for $n<0$. From the theoretical point of view such a 
situation seems to be unsatisfactory because in the space of all dynamics 
systems on the plane they form set of zero measure (the Peixoto theorem).

The advantage of representing dynamics in terms of Hamiltonian is to 
discuss how trajectories with interesting properties are distributed 
on phase plane.

\acknowledgments
M. Szyd{\l}owski acknowledges the support of 2002/2003 Jagiellonian 
University Rector's Scholarship.

\end{document}